\documentclass[lettersize, journal]{IEEEtran}
\usepackage{float}
\usepackage{caption}
\usepackage{ntheorem}
\usepackage{color}
\usepackage{amsmath}  
\usepackage{array}    
\usepackage{cite}
\usepackage{amsmath,amssymb,amsfonts}
\usepackage{graphicx}
\usepackage{textcomp}
\usepackage{xcolor}
\usepackage{svg}
\usepackage{subfigure}
\usepackage{mathtools}
\usepackage{epsfig}

\usepackage[ruled,vlined]{algorithm2e}
\usepackage{algpseudocode}
\usepackage{booktabs} 

\newtheorem{remark}{Remark}
\newtheorem{theorem}{Theorem}

\newtheorem{lemma}{Lemma}

\newtheorem{corollary}{Corollary}

\allowdisplaybreaks
\makeatletter
\def\ScaleIfNeeded{%
\ifdim\Gin@nat@width>\linewidth \linewidth \else \Gin@nat@width
\fi } \makeatother

\graphicspath{{Fig/}}

\begin{document}

\title{Interference-Robust Broadband Rapidly-Varying MIMO Communications: A Knowledge-Data Dual Driven Framework}
\author{Jingjing Zhao, Jing Su,  Kaiquan Cai, Yanbo Zhu, \\
Yuanwei Liu,~\IEEEmembership{Fellow,~IEEE},
and Naofal Al-Dhahir,~\IEEEmembership{Fellow,~IEEE}
\thanks{J. Zhao, J. Su, K. Cai, and Y. Zhu are with the Beihang University, Beijing,
China, and also with the State Key Laboratory of CNS/ATM, Beijing, China. (email:\{jingjingzhao, jingsu, caikq, zhuyanbo\}@buaa.edu.cn). }
\thanks{Y. Liu is with the Queen Mary University of London, London, UK. (email: yuanwei.liu@qmul.ac.uk).}
\thanks{N. Al-Dhahir is with the University of Texas at Dallas, Richardson, USA. (email: aldhahir@utdallas.edu).}
}

\maketitle
\begin{abstract}
A novel time-efficient framework is proposed for improving the robustness of a broadband multiple-input multiple-output (MIMO) system against unknown interference under rapidly-varying channels. A mean-squared error (MSE) minimization problem is formulated by optimizing the beamformers employed. Since the unknown interference statistics are the premise for solving the formulated problem, an interference statistics tracking (IST) module is first designed. The IST module exploits both the time- and spatial-domain correlations of the interference-plus-noise (IPN) covariance for the future predictions with data training. Compared to the conventional signal-free space sampling approach, the IST module can realize zero-pilot and low-latency estimation. Subsequently, an interference-resistant hybrid beamforming (IR-HBF) module is presented, which incorporates both the \textit{prior} knowledge of the theoretical optimization method as well as the data-fed training. Taking advantage of the interpretable network structure, the IR-HBF module enables the simplified mapping from the interference statistics to the beamforming weights. The simulations are executed in high-mobility scenarios, where the numerical results unveil that: 1) the proposed IST module attains promising prediction accuracy compared to the conventional counterparts under different snapshot sampling errors; and 2) the proposed IR-HBF module achieves lower MSE with significantly reduced computational complexity.

\end{abstract}
\begin{IEEEkeywords}
Hybrid beamforming, interference tracking, interference resistance, knowledge-data dual driven network, multiple-input multiple-output.
\end{IEEEkeywords}
\section{Introduction}
Due to the broadcasting nature of the wireless medium, communication systems are vulnerable to various forms of interference, such as co-channel/adjacent-channel interference caused by legacy systems on the same/adjacent channels~\cite{ITU2020}. 
The interference can lead to significant reductions of the spectral efficiency and reliability of the legitimate transmissions, which implies that the physical layer should cope with the distortion when focusing on the single-link aspect. Multiple-input multiple-output (MIMO) techniques have gained significant interest in recent years, for adaptively protecting communications systems from interference by leveraging highly directional beams~\cite{9024145}. Specifically, MIMO physical layer processing against interference normally involves two steps. The first step is the estimation of the unknown interference characteristics (such as the covariance matrix), given that the interfering systems are normally non-cooperative~\cite{7345553}. The second step is interference-resistant beamforming for simultaneously mitigating the interference and enhancing the desired signal strength~\cite{8335273}. 

Despite the effectiveness of the MIMO technique in resisting interference by exploiting the degrees of freedom (DoF) in the spatial domain, interference-robust MIMO processing is still a challenging issue under rapidly-varying communications channels\footnote{One typical example of the interfering use case under rapidly-varying channels is the L-band digital aeronautical communications system (LDACS) embedded between the distance measurement equipment (DME) system in the $960$~MHz-$1164$~MHz frequency band~\cite{10024899}. Due to the out-of-band radiation of the random DME impulses, the signal-to-interference power ratio at the LDACS receiver can reach to $-3.8$~dB under the typical en-route scenario~\cite{9535223}.}, which is mainly caused by the following two main reasons. Firstly, snapshot sampling for the interference observation normally requires the projection onto a signal-free space to avoid mixture with the desired signals, which can cause unaffordable pilot overhead, thereby degrading communication efficiency. Secondly, the beamforming matrices need to be updated each time the communication channels change (as dictated by the coherence time), which requires that the interference-resistant beamforming should operate with a relatively low computational complexity. Therefore, it is of vital importance for designing time-efficient interference-robust MIMO schemes to increase the transmission reliability for rapidly-varying communication scenarios.
\vspace{-0.2in}
\subsection{Related Works}
\subsubsection{Studies on Interference Estimation in MIMO Systems}
One common theme of existing interference estimation methods is to process some specific subset of the received samples in the time domain, frequency domain, or spatial domain to avoid mixture with the desired signals. For example, the authors of~\cite{8094873} estimated the interference power with several intentionally unused pilot sequences, and exploited the generalized likelihood ratio test for detecting the presence of the interference. The authors of~\cite{6666096} utilized random training for the detection of the pilot contamination attack caused by the interference in the pilot phase. Moreover, an energy ratio detector was proposed in~\cite{EDR}, which exploited the asymmetry of signal power levels at the legitimate receiver under a pilot spoofing attack. The authors of~\cite{9527146} deployed unused pilot sequences which span the received signals and could be projected to estimate the interference subspace. Considering that the active interference caused by smart jammers could evade mitigation by transmitting only at specific instants, a unified interference estimation and mitigation scheme was proposed in~\cite{10048559}. To be more specific, for interference suppression problems aiming at signal-to-interference-plus-noise ratio (SINR) maximization or mean square error (MSE) minimization, the interference-plus-noise (IPN) covariance matrix is a key factor that should be obtained in advance. To this end, an IPN estimation algorithm was proposed in~\cite{6180020}, which uses the spatial spectrum distribution, and projects the received signal onto the spatial space to exclude the desired signals. Furthermore, a spatial power spectrum sampling method was proposed in~\cite{7345553} to reconstruct the IPN covariance matrix more efficiently. However, for both cases in~\cite{6180020,7345553}, the IPN covariance matrix estimation requires high pilot overhead and computational complexity due to the large number of samples involved in both spectrum estimation and matrix multiplication. Moreover, the spatial subspace projection relies on the presumed steering vectors, which is challenging to obtain especially under pilot contamination.   
\subsubsection{Studies on Interference Suppression in MIMO Systems}
Building upon research into interference estimation, some works have focused on interference suppression in MIMO communications systems~\cite{8781945,8705374,10422749}. The authors of~\cite{8781945} designed the analog combiners of the receivers to cancel the strong interference outside the cell by creating an orthogonal component to the interfering channel. In~\cite{8705374}, a joint transceiver anti-jamming optimization scheme was proposed to minimize the power consumption in the uplink MIMO system. The authors of~\cite{10422749} proposed an anti-interference hybrid beamforming (HBF) algorithm based on the matrix decomposition-based alternating minimization method. Nevertheless, the works~\cite{8781945,8705374} designed the anti-interference schemes with the premise that the interfering channel state information (CSI) was known. In addition, the theoretical optimization-based interference-resistant HBF algorithms proposed in~\cite{8781945,8705374,10422749} suffer from high computational complexity, which makes them unsuitable for the rapidly-varying channels. 

As a computationally-efficient substitute for the conventional optimization model-based approach, the data-driven deep learning (DL) has recently attracted significant attention for solving complex wireless communication problems with low latency~\cite{8054694}. 
With recent advancements in the field, researchers have explored DL-based beamforming design in MIMO systems~\cite{9763852}.
However, as the network structure is treated as a ``black-box", purely data-driven DL has the drawbacks of poor interpretability and high memory requirements. Therefore, the concept of \textit{knowledge-data dual driven (KDDD)} networks has been considered in the present context. Through incorporating the available \textit{prior} knowledge of the theoretical model into the network design, the KDDD approach is capable of simplifying the training process with a reduction in training parameters. 
Some studies have attempted to apply KDDD networks in MIMO beamforming design~\cite{8715338,10185621,10286447}. Specifically, {the authors of~\cite{8715338} discussed model-driven DL approaches that combined the communication domain knowledge with DL to reduce the demand for computing resources and training time.}
The authors of~\cite{10185621} combined manifold optimization (MO) and deep neural networks (NNs) to design the hybrid precoder with the aim of spectral efficiency maximization.
Additionally, the authors of~\cite{10286447} designed the analog beamformer based on the unfolded projected gradient descent algorithm. 
To the best of the authors' knowledge, the KDDD HBF design for MIMO systems with the existence of interference has not yet been well studies. 

\vspace{-0.1in}
\subsection{Motivations and Contributions}
Despite the aforementioned progress on interference estimation and suppression in MIMO systems, the existing schemes commonly use excessive pilot overhead and suffer from high computational complexity, making them not applicable to rapidly-varying communications scenarios. In this work, we focus on the time-efficient interference-robust MIMO communications framework design given unknown interference. Aiming at MSE minimization, our proposed framework contains an interference statistics tracking (IST) module and an interference-resistant hybrid beamforming (IR-HBF) module. For the sake of brevity, the main contributions are summed up as follows.
\begin{itemize}
 \item We develop a novel IST module that predicts the IPN covariance matrices in parallel with data training. Compared to the existing model-based approaches that require signal-free space projection, the proposed scheme can efficiently reduce the computational complexity and pilot overhead. Particularly, the parallel prediction mechanism is applied to establish the mapping between the historical IPN covariance matrices and the future ones. By exploiting both the time- and spatial-domain correlations, the convolutional operations are executed for improving the IPN covariance accuracy against initial sampling errors.
 \item We propose a novel low-complexity IR-HBF module, which possesses both the interpretability and the robustness against imperfect interference statistics. By deploying the \textit{prior knowledge} of alternating optimization (AO) operations, a KDDD interference-resistant network (KDDD-IRN) is designed with simpler processing and fewer trainable parameters compared to the ``black-box" counterpart. Moreover, unlike the traditional AO method requiring unaffordable number of iterations until convergence, the KDDD-IRN fixes the number of iterations based on the complexity and latency constraints, through incorporating the trainable gradient descent step sizes. 

\item We show that the proposed IST module can achieve promising IPN covariance prediction accuracy under different sampling errors, compared to the existing data-driven benchmarks. We also demonstrate that the proposed IR-HBF module outperforms the conventional iteration-based optimization approach under the same number of iterations and shows robustness against predicted IPN errors.
\end{itemize}
\subsection{Organization and Notation}
The rest of this paper is organized as follows.
In Section II, the system model is first introduced, which is followed by the MSE minimization problem formulation. In Section III, the IST module for the IPN covariance matrices prediction is introduced. In Section IV, the IR-HBF module with low complexity and high interpretability is further elaborated. The simulation results are presented in Section V and the conclusions are drawn in Section VI.

$\textit {Notation}$: Scalars, vectors and matrices are denoted by italic letters, bold-face lower-case, and bold-face upper-case, respectively. $\mathbb{C}^{N\times M}$ denotes the set of $N\times M$ complex-valued matrices. Superscripts $(\cdot)^*, (\cdot)^T, (\cdot)^H, (\cdot)^{-1}$ denote the conjugate, transpose, conjugate transpose, and inversion operators, respectively. $|\cdot|$,  $\left\|\cdot\right\|$ denote the  determinant and Euclidean norm of a matrix. $\text{Tr}\left(\cdot\right)$, $\left\|\cdot\right\|_F$ and $\text{vec}\left(\cdot\right)$ denote the trace, Frobenius norm and vectorization of a matrix,  respectively. $[\cdot]_{m,n}$ denotes the $(m,n)$-th element of a matrix.
$\mathbb{E}$ denotes the expectation operator. All random variables are assumed to be \textit{zero} mean. The main symbols in this papers are summarized in Table~\ref{tab:parameters}.
\renewcommand{\arraystretch}{1.2}
\begin{table}[h]
\centering
\captionsetup{labelformat=empty}  
\caption{\textnormal{TABLE I}\\\textsc{List of Symbols}}    
\label{tab:parameters}
{
\begin{tabular}{ >{\centering\arraybackslash}m{1.5cm} | >{\centering\arraybackslash}m{6cm}}
\hline
\textbf{Symbol} & \textbf{Definition} \\
\hline
\hline
$X$ & Number of subcarriers \\
\hline
$N_{\text{s}}$  & Number of data streams\\
\hline
$K_\text{A}$, $K_\text{B}$ & Number of antennas at the AC and LGS\\
\hline
$K_{\text{RF}}^{\text{A}}$, $K_{\text{RF}}^{\text{B}}$ & Number of RF chains at the AC and LGS\\
\hline
$\mathbf{s}_{x}^t$ & LGS transmitted signal on subcarrier $x$ in frame $t$\\
\hline
$\mathbf{y}_{x}^t$& AC received signal on subcarrier $x$ in frame $t$\\
\hline
$\mathbf{V}_{\text{BB},x}$& LGS digital precoder matrix on subcarrier $x$\\
\hline
$\mathbf{V}_{\text{RF}}$& LGS analog precoder matrix\\
\hline
$\mathbf{W}_{\text{BB},x}$ & AC digital combiner matrix on subcarrier $x$\\
\hline
$\mathbf{W}_{\text{RF}}$& AC analog combiner matrix\\
\hline
$\mathbf{V}_{x}$& LGS hybrid precoder matrix on subcarrier $x$\\
\hline
$\mathbf{W}_{x}$& AC hybrid combiner matrix on subcarrier $x$\\
\hline
$\mathbf{H}_{\text{BA}}^{x}$& LGS-AC channel matrix on subcarrier $x$\\
\hline
${\mathbf{R}}_x$& IPN covariance matrix on subcarrier $x$\\
\hline
${\beta_{x}}$& Scaling factor on subcarrier $x$\\
\hline
$\mathbf{i}_{x}$& Received interfering signal on subcarrier $x$\\
\hline
$\mathbf{n}_x$& Additive white Gaussian noise on subcarrier $x$\\
\hline
$\mathbf{d}_x$& Interference-plus-noise signal on subcarrier $x$\\
\hline
${S}$ & Number of sampling snapshots\\
\hline
$U$& Number of NLoS paths\\
\hline
$\mathbf{R}^t$, $\hat{\mathbf{R}}^t$ & Actual and estimated IPN covariance matrices\\
\hline
$\mathbf{E}^t$& Estimation error of ${\mathbf{R}}^t$\\
\hline
$\mathbf{R}_{\text{(e)}}$, $\mathbf{R}_{\text{(d)}}$& IPN covariance matrix input to encoder and decoder\\
\hline
$P$& Number of frames for historical IPN observation\\
\hline
$L$& Number of frames for future IPN prediction\\
\hline
$\boldsymbol{\gamma}_{\text{B}}^{i},\boldsymbol{\gamma}_{\text{A}}^{i}$& Trainable parameters in KDDD-IRN\\
\hline
\end{tabular}
}
\end{table}
\section{System Model and Problem Formulation}
In this section, we first introduce the broadband MIMO communications system under rapidly-varying channels in the presence of interference, and then formulate a MSE minimization problem in response to reliable communications requirements.
\begin{figure}[H]
\centering
\includegraphics[scale=0.35]{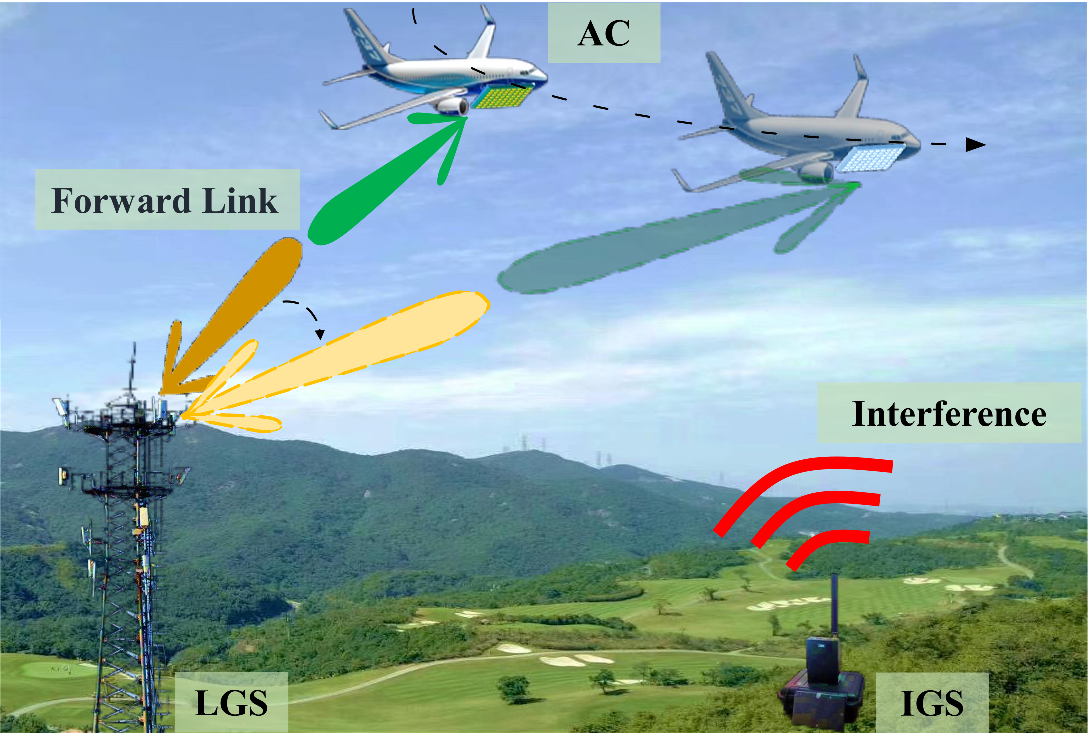}
	\caption{Forward link transmission of an A2G MIMO system under random interference attack.}
    \label{FIG_1}
\end{figure}

\subsection{System Model}
We consider the forward link transmission of an air to ground (A2G) broadband MIMO system as an example of the communication use case with rapidly-varying channels, which consists of one $K_{\text{B}}$-antenna legitimate ground station (LGS) and one $K_{\text{A}}$-antenna aircraft (AC) in the presence of a single-antenna interfering ground station (IGS), as depicted in Fig.~\ref{FIG_1}. Note that the IGS is assumed to be a non-cooperative node with an unknown antenna configuration. Nevertheless, even if the IGS is deployed with an antenna array, it can still be equivalently treated as a single-antenna source after beamforming.
The LGS and AC are both equipped with a uniform planar array (UPA) formed antenna elements and the hybrid analog-digital beamforming architecture for cost saving.   
The orthogonal frequency division multiplexing (OFDM) communication scheme is adopted at the LGS to transmit independent data streams on $X$ subcarriers. The signal vector transmitted on the $x$-th subcarrier is denoted by $\mathbf{s}_x\in\mathbb{C}^{N_\text{s}}$ which satisfies $\mathbb{E}\left[\mathbf{s}_x\mathbf{s}_x^{H}\right] = \mathbf{I}_{N_\text{s}}$. To guarantee the effectiveness of the communication carried by the limited number of radio frequency (RF) chains, the number of the transmitted data streams on each subcarrier is constrained by $N_{\text{s}}\leq \min\left\{K_{\text{RF}}^{\text{B}}, K_{\text{RF}}^{\text{A}}\right\}$, where $K_{\text{RF}}^{\text{B}}$ and $K_{\text{RF}}^{\text{A}}$ represent the number of RF chains at the LGS and AC, respectively. For generality, the IGS is assumed to generate random interfering signals with high power, which is modeled by narrow and random pulses of duration approximately equal to the inverse of the signal bandwidth, i.e., the time domain sampling interval of the OFDM symbol. Thus, without interference suppression, the impulses in the time domain corrupt all the subcarriers of an OFDM symbol, which yields a burst of very noisy frequency domain symbols that can significantly degrade the communication reliability.

In the investigated system, the time resource is divided into multiple frames, where the channel condition is assumed to be unchanged within each frame. Each frame contains three phases, i.e., interference estimation, channel estimation, and data transmission phases. For the data transmission phase in the $t$-th frame, the received signal at the AC over the $x$-th subcarrier is given as 
\begin{align}
     \mathbf{y}_{x}^t=&(\mathbf{W}_{\text{BB},x}^t)^{H}(\mathbf{W}_{\text{RF}}^t)^{H}\mathbf{H}_{\text{BA}}^{x,t}\mathbf{V}_{\text{RF}}^t\mathbf{V}_{\text{BB},x}^t\mathbf{s}_{x}^t\nonumber\\
    & + (\mathbf{W}_{\text{BB},x}^t)^{H}(\mathbf{W}_{\text{RF}}^t)^{H}\mathbf{i}_{x}^t+(\mathbf{W}_{\text{BB},x}^t)^{H}(\mathbf{W}_{\text{RF}}^t)^{H}\mathbf{n}_{x}^t,
    \label{received signal}
\end{align}
where $\mathbf{V}_{\text{BB},x}^t\in \mathbb{C}^{K_{\text{RF}}^{\text{B}}\times N_{\text{s}}}$ and $\mathbf{V}_{\text{RF}}^t\in \mathbb{C}^{K_{\text{B}}\times K_{\text{RF}}^{\text{B}}}$ represent the digital and analog precoder matrices at the LGS, respectively, $\mathbf{W}_{\text{BB},x}^t\in \mathbb{C}^{K_{\text{RF}}^{\text{A}}\times N_{\text{s}}}$ and $\mathbf{W}_{\text{RF}}^t\in \mathbb{C}^{ K_{\text{A}}\times K_{\text{RF}}^{\text{A}}}$ denote the digital and analog combiner matrices at the AC, respectively, and $\mathbf{H}_{\text{BA}}^{x,t}\in \mathbb{C}^{K_{\text{A}}\times K_{\text{B}}}$ is the LGS-AC channel matrix. Moreover, $\mathbf{s}_x^t\in\mathbb{C}^{N_{\text{s}}}$, $\mathbf{i}_{x}^t\in\mathbb{C}^{K_{\text{A}}}$, and $\mathbf{n}_x^t\in\mathbb{C}^{K_{\text{A}}}$ represent the transmitted signal at the LGS, the received interfering signal at the AC, and the additive white Gaussian noise (AWGN) with zero mean and a covariance matrix of $\sigma_n^{2}\mathbf{I}_{K_{\text{A}}}$, respectively. Note that $\mathbf{s}_{x}^t$, $\mathbf{i}_{x}^t$ and $\mathbf{n}_{x}^t$ are statistically independent.
Considering the limited scatterers in the A2G propagation environment, we adopt the geometric Saleh-Valenzuela channel model~\cite{6717211} with one LoS path and $U$ NLoS paths for representing the LGS-AC channel matrix $\mathbf{H}_{\text{BA}}^{x,t}$, which can be expressed as
\begin{align}
    \mathbf{H}_{\text{BA}}^{x,t} & = \alpha_{x,\text{LoS}}^{t}{\mathbf{a}_{\text{A}}}\left(\phi_{\text{A,LoS}}^{\text{azi},t}, \phi_{\text{A,LoS}}^{\text{ele},t}\right){\mathbf{a}_\text{B}^H}\left(\phi_{\text{B,LoS}}^{\text{azi},t}, \phi_{\text{B,LoS}}^{\text{ele},t}\right)\nonumber\\
    & + \sqrt{\frac{1}{U\rho_{\text{R}}}}\sum_{u=1}^U{\alpha_{x,u}^{t}}{{\mathbf{a}_{\text{A}}}\left(\phi_{\text{A},u}^{\text{azi},t}, \phi_{\text{A},u}^{\text{ele},t}\right){\mathbf{a}_\text{B}^H}\left(\phi_{\text{B},u}^{\text{azi},t}, \phi_{\text{B},u}^{\text{ele},t}\right)},
    \label{eq:channel-model}
\end{align}
where $\rho_\text{R}$ denotes the Rician factor, $\phi _{\text{A},\cdot}^{\text{azi},t}$ ($\phi _{\text{A},\cdot}^{\text{ele}, t}$) and $\phi_{\text{B},\cdot}^{\text{azi}, t}$ ($\phi_{\text{B},\cdot}^{\text{ele}, t}$) denote the azimuth (elevation) angle of arrival (AoA), and the azimuth (elevation) angle of departure (AoD) of the channel paths. ${\mathbf{a}_{\text{A}}}\left(\phi_{\text{A},\cdot}^{\text{azi},t}, \phi_{\text{A},\cdot}^{\text{ele},t}\right)\in \mathbb{C}^{K_\text{A}\times 1}$ and ${\mathbf{a}_\text{B}^H}\left(\phi_{\text{B},\cdot}^{\text{azi},t}, \phi_{\text{B},\cdot}^{\text{ele},t}\right)\in \mathbb{C}^{1\times K_{\text{B}}}$ denote the array response vector at the AC and LGS, respectively. $\alpha_{x,\text{LoS}}^{t}$ ($\alpha_{x,u}^{t}$) represents the complex-valued gain of the LoS (NLoS) path. $\alpha_{x,u}^{t}$ is expressed as
\begin{align}
& \alpha_{x,u}^{t} = {h_u^t}e^{-j2\pi\frac{ x\tau_u^t}{T_sX}},
\end{align}
where $h_u^t$ and $\tau_u^t$ denote the large-scale fading gain and the delay of the $u$-th NLoS in the $t$-th frame, respectively. $\alpha_{x,\text{LoS}}^{t}$ is found by substituting $h_u^t$ and $\tau_u$ with $h_{\text{LoS}}^t$ and $\tau_{\text{LoS}}^t$, respectively. 
\subsection{Problem Formulation}
In this work, we aim at minimizing the MSE between the received signals and the transmitted desired signals to characterize the transmission reliability. As the MSE in each data frame follows a similar expression, we omit the frame index $t$ and rewrite ${\mathbf{H}}_{\text{BA}}^{x,t}$, $\mathbf{V}_{\text{BB},x}^{t}$, $\mathbf{V}_{\text{RF}}^t$, $\mathbf{W}_{\text{BB},x}^t$, and $\mathbf{W}_{\text{RF}}^{t}$ as ${\mathbf{H}}_{\text{BA}}^{x}$, $\mathbf{V}_{\text{BB},x}$, $\mathbf{V}_{\text{RF}}$, $\mathbf{W}_{\text{BB},x}$, and $\mathbf{W}_{\text{RF}}$, respectively, in the rest of this section. The MSE is given by
\begin{equation}
\label{eq:MSE1}
\begin{aligned}
{\text{MSE}[x]}
& =\mathbb{E}\left(\left\|\mathbf{s}_{x}-\beta_{x}^{-1}\mathbf{y}_{x}\right\|^{2}\right),
\end{aligned}
\end{equation}
where $\beta_x$ is the scaling factor for the gain of the filter chain $\mathbf{W}_x^H\mathbf{H}_{\text{BA}}^x\mathbf{V}_x$ with the given transmit power and the noise power. By using the properties 
$\left\|\mathbf{a}\right\|^{2}=\mathrm{Tr}(\mathbf{a}\mathbf{a}^H)$ and $\mathbb{E}\left[\mathrm{Tr}\left(\mathbf{a}\mathbf{a}^H\right)\right]=\mathrm{Tr}\left[\mathbb{E}\left(\mathbf{a}\mathbf{a}^H\right)\right]$ for a column vector $\mathbf{a}$ , we can further derive the MSE as
\begin{equation}
\label{eq:MSE}
\begin{aligned}
{\text{MSE}[x]}
&=\mathrm{Tr}\left\{\mathbb{E}\left[\left(\mathbf{s}_{x}-\beta_{x}^{-1}\mathbf{y}_{x}\right)\left(\mathbf{s}_{x}-\beta_{x}^{-1}\mathbf{y}_{x}\right)^H\right]\right\} \\
&=\mathrm{Tr}( 
\beta_x^{-2}\mathbf{W}_x^H\mathbf{H}_{\text{BA}}^x\mathbf{V}_x\mathbf{V}_x^H(\mathbf{H}_{\text{BA}}^x)^H\mathbf{W}_x\\
&\ \ \ \ -\beta_x^{-1}\mathbf{W}_x^H\mathbf{H}_{\text{BA}}^x\mathbf{V}_x-\beta_{x}^{-1}\mathbf{V}_{x}^{H}(\mathbf{H}_{\text{BA}}^{x})^{H}\mathbf{W}_{x}\\
&\ \ \ \ +\beta_{x}^{-2}\mathbf{W}_{x}^{H}\mathbf{R}_{x}\mathbf{W}_{x}+\mathbf{I}_{N_{s}}),
\end{aligned}
\end{equation}
where $\mathbf{V}_x \triangleq \mathbf{V}_{\text{RF}}\mathbf{V}_{\text{BB},x}$ and $\mathbf{W}_x \triangleq \mathbf{W}_{\text{RF}}\mathbf{W}_{\text{BB},x}$ are the joint precoder and combiner at the LGS and AC, respectively, and $\mathbf{R}_x = \mathbb{E}(\mathbf{d}_x\mathbf{d}_x^H)\in\mathbb{C}^{K_{\text{A}}\times {K_{\text{A}}}}$ denotes the covariance matrix of the received IPN over the AC antennas, with $\mathbf{d}_x=\mathbf{i}_{x}+\mathbf{n}_{x}$. The MSE minimization problem is formulated as follows
\begin{subequations}
\label{eq:optimization_problem}
\begin{equation}
\label{eq:6aMSE_objective}
\min_{\mathbf{W}_\text{RF},\mathbf{W}_{{\text{BB},x}},\mathbf{V}_\text{RF},\mathbf{V}_{{\text{BB},x}},\beta_{x}}
\sum_{x=0}^{X-1}\text{MSE}[{x}],
\end{equation}
\begin{equation}
\label{eq:6bpower_constraint}
    {\rm{s.t.}} \ \ \text{Tr}\left(\mathbf{V}_{\text{RF}}\mathbf{V}_{\text{BB},x}\left(\mathbf{V}_{\text{BB},x}\right)^{H}\mathbf{V}_{\text{RF}}^{H}\right)\leq 1, \forall x,
\end{equation}
\begin{equation}
\label{eq:6cA_analog_constraint}
|[\mathbf{V_{\text{RF}}}]_{i, j}|=1,\forall i,j,
\end{equation}
\begin{equation}
\label{eq:6dB_analog_constraint}
|[\mathbf{W_{\text{RF}}}]_{m, l}|=1,\forall m,l,
\end{equation}
\end{subequations}
where \eqref{eq:6bpower_constraint} is the normalized transmit power constraint, while \eqref{eq:6cA_analog_constraint} and \eqref{eq:6dB_analog_constraint} are the constant modulus constraints of the RF precoder and RF combiner, respectively. 

With the derivation of the MSE given in~(\ref{eq:MSE}), one can see that the above optimization problem can only be solved assuming known $\mathbf{R}_x$ and ${\mathbf{H}}_{\text{BA}}^{x}$. Since there are many methods present in the literature for channel estimation in MIMO-OFDM systems (e.g., \cite{10464206}), here we do not deal with the channel estimation and focus specifically on the effect of random interference given the page limit. It is apparent that the snapshot sampling for the observation of $\mathbf{R}_x$ will require significant pilot overhead under fast time-varying channels, thereby degrading the communication efficiency. Moreover, as the optimization problem~(\ref{eq:optimization_problem}) should be solved once $\mathbf{R}_x$ and ${\mathbf{H}}_{\text{BA}}^{x}$ changes (dictated by the coherence duration of the A2G channels), our proposed solving approach should operate with relatively low computational complexity. Unfortunately, (\ref{eq:optimization_problem})
is intractable due to the highly-coupled optimization variables as well as the constant modulus constraints, while the existing iteration-based optimizers~\cite{7397861} take a large amount of time-executing iterations until convergence. To resolve the above two challenges, we first propose an IST strategy in Section III, which leverages the set of past observations and predicts the IPN covariance matrices in future frames with low pilot overhead and high robustness. Then, in Section IV, with the acquired $\mathbf{R}_x, x\in\left\{1, ..., X\right\}$, we carry out the low-complexity IR-HBF optimization by getting access to the previously encountered or simulated channel realizations to achieve satisfactory performance within a predefined number of iterations.
\section{Interference Statistics Tracking}
In this section, we propose a data-driven IST module to aid the subsequent IR-HBF module in Section IV. Note that the efficient IST relies on historical observations, while the key for the initial IPN observation is to project the interference onto a signal-free space so that the observed IPN is not mixed with the desired signal. In the current literature, such as~\cite{6742716}, null subcarriers are deployed in each OFDM symbol during the interference estimation phase for IPN estimation. Since the interference projection is rank-deficient as the projection subspace is smaller than the total subcarriers space, methods like interpolation or compressed sensing~\cite{6742716} can be applied to recover the full-space observation. Due to the page limit, in this work, we focus on the IPN prediction problem assuming that the initial estimation results have been acquired. 
Let $\mathbf{d}_x^t[s]$ denote the $s$-th snapshot sampling of the IPN in the $t$-th frame. Then, the theoretical IPN covariance matrix ${\mathbf{R}}_x^t$ at the $x$-th subcarrier can be approximated by the empirical average as 
\begin{equation}
\hat{\mathbf{R}}_x^t =\frac{1}{S}\sum_{s=1}^{S}\mathbf{d}_x^t[s]\left(\mathbf{d}_x^{t}[s]\right)^H,
\label{eq:cov_cal}
\end{equation}
 where ${S}$ is total number of sampling snapshots. As $S$ increases, $\hat{\mathbf{R}}_x^t$ converges to the theoretical covariance matrix ${\mathbf{R}}_x^t$ under stationary and ergodic assumptions. When $S$ is small, there is a large gap between $\hat{\mathbf{R}}_x^t$ and ${\mathbf{R}}_x^t$, which dramatically affects the subsequent interference suppression performance. However, to guarantee the performance, the null-subcarriers allocation within the OFDM symbol apparently wastes efficient communication bandwidth especially under the rapidly-varying channels. 

As the IPN covariance matrices in each data frame are temporally correlated, an intuitive idea for replacing the IPN estimation approach above is to predict the future IPN covariance based on the historical estimation results. In fact, although solving the IPN tracking problems is lacking in the current literature, some data-driven techniques have been proposed for the prediction of time-varying parameters in wireless communications, such as the communication channels~\cite{9676455,9044427}. Specifically, the NNs are trained with the set of past observations to learn the changes of the predicted parameters. With the hand-tuned network training weights, the future parameter values can be predicted rapidly even with zero pilot overhead. However, the conventional NN frameworks, such as long short-term memory (LSTM)~\cite{9676455} and recurrent neural network (RNN)~\cite{9044427}, work in a sequential manner and lead to the propagation of prediction errors, thereby dramatically degrading the prediction accuracy with the evolution of data frames. In this work, we propose a novel IST module that deploys a parallel vector mapping network named 3DConvTransformer. The 3DConvTransformer network contains a three-dimensional convolutional neural network (3D-CNN) cascaded by a transformer~\cite{9832933}, as shown in Fig.~\ref{3DConvTransformer}. Considering that the statistical expectation approximation in~(\ref{eq:cov_cal}) causes inevitable estimation errors, the historical IPN covariance matrices are first fed into a 3D-CNN to mitigate the sampling errors. Note that compared to the deterministic signal processing algorithms, the 3D-CNN can flexibly capture the complicated time- and spatial-domain correlations of the IPN covariance matrices in a data-driven approach.
Then, the subsequent transformer deploys the refined $P$ IPN covariance matrices output from the 3D-CNN to predict the future ones in the next $L$ successive frames in parallel. In the following, the 3DConvTransformer network is first explained, followed by the computational complexity analysis.
\subsection{3DConvTransformer-Based IST}
In the $t_0$-th frame, suppose that the AC receiver has collected the IPN covariance matrices from the past $P$ frames, i.e., $\left\{\mathbf {\hat{R}}^{(t_0-P+1)}, ..., \mathbf{\hat{R}}^{t_0}\right\}$ with $\left\{\hat{\mathbf{R}}^{t}\right\}_{\left[K_{\text{A}}, K_{\text{A}},x\right]} = \hat{\mathbf{R}}_x^t, \forall x\in\left\{1, ..., X\right\}$. To simplify the presentation, we also define the actual IPN covariance matrices involving the subcarrier dimension as $\left\{{\mathbf{R}}^{t}\right\}_{\left[K_{\text{A}}, K_{\text{A}},x\right]} = {\mathbf{R}}_x^t, \forall x\in\left\{1, ..., X\right\}$. Denote the estimation error of ${\mathbf{R}}^t, t\in\left\{t_0-P+1, ..., t_0\right\}$ by $\mathbf{E}^t\in\mathbb{C}^{K_{\text{A}}\times {K_{\text{A}}}\times X}$ with each element modeled by a random variable following the Gaussian distribution of mean $0$ and variance $\sigma_e^2$, then we have
\begin{equation}
 \mathbf{R}^t=\hat{\mathbf{R}}^t + \mathbf{E}^t, t\in\left\{t_0-P+1, ..., t_0\right\}.
\end{equation}
Since $\hat{\mathbf{R}}^t$ possesses the {time-domain correlation} brought by the time-varying channels and elements in $\hat{\mathbf{R}}_x^t$ are correlated in the antenna domain through the steering vectors, we propose to utilize the CNN network to first sanitize $\hat{\mathbf{R}}^t$ by leveraging the \textit{bidimensional correlations} through convolution operations~\cite{9209095}\cite{9127834}. Let $\left\{\dot{{\mathbf{{R}}}}^{(t_0-P+1)},...,\dot{{\mathbf{{R}}}}^{t_0}\right\}$ denote the $P$-length output of the CNN, which is the mapping from the historical observations and the training parameters set $\Theta_{\text{conv}}$ with the function $\mathcal{F}_{\text{conv}}$ as follows
\begin{equation}
\left\{\dot{{\mathbf{{R}}}}^{(t_0-P+1)},...,\dot{{\mathbf{{R}}}}^{t_0}\right\} = \mathcal{F}_{\text{conv}}\left(\left\{\mathbf{{R}}^{(t_0-P+1)},...,\mathbf{{R}}^{t_0}\right\}, \Theta_{\text{conv}}\right).
\end{equation}

Since the input $\left\{\mathbf {\hat{R}}^{(t_0-P+1)},..., \mathbf {\hat{R}}^{t_0}\right\}$ to the CNN is a four-dimensional matrix with size of $ K_{\text{A}}\times K_{\text{A}} \times X \times P$, the 3D convolutional kernels are required to fully process the input without dimension reduction operations and avoid information loss of the original input. To simplify the presentation, we also define the input IPN covariance matrices involving the time dimension as $\mathbf{\hat{R}}_{\left[K_{\text{A}},K_{\text{A}},X,t\right]} = {\mathbf{\hat{R}}}^t, \forall t\in\left\{t_0-P+1, ..., t_0\right\}$. It is known that NNs are more effective for real-valued operations than complex-valued operations. Therefore, the real and imaginary parts of the IPN covariance matrix $\mathbf{\hat{R}}$ are extracted and reshaped into a four-dimensional real-valued matrix $\mathbf {\tilde{R}}$ as follows
\begin{equation}\begin{cases}\tilde{\mathbf{R}}_{[1:K_\mathrm{A}, K_\mathrm{A}, X, P]}=\mathcal{R}\left\{\hat{\mathbf{R}}\right\},\\\tilde{\mathbf{R}}_{[K_\mathrm{A}+1:2K_\mathrm{A}, K_\mathrm{A}, X, P]}=\mathcal{I}\left\{\hat{\mathbf{R}}\right\}.\end{cases}\end{equation}

\begin{figure}
	\centering
	\includegraphics[scale=0.5]{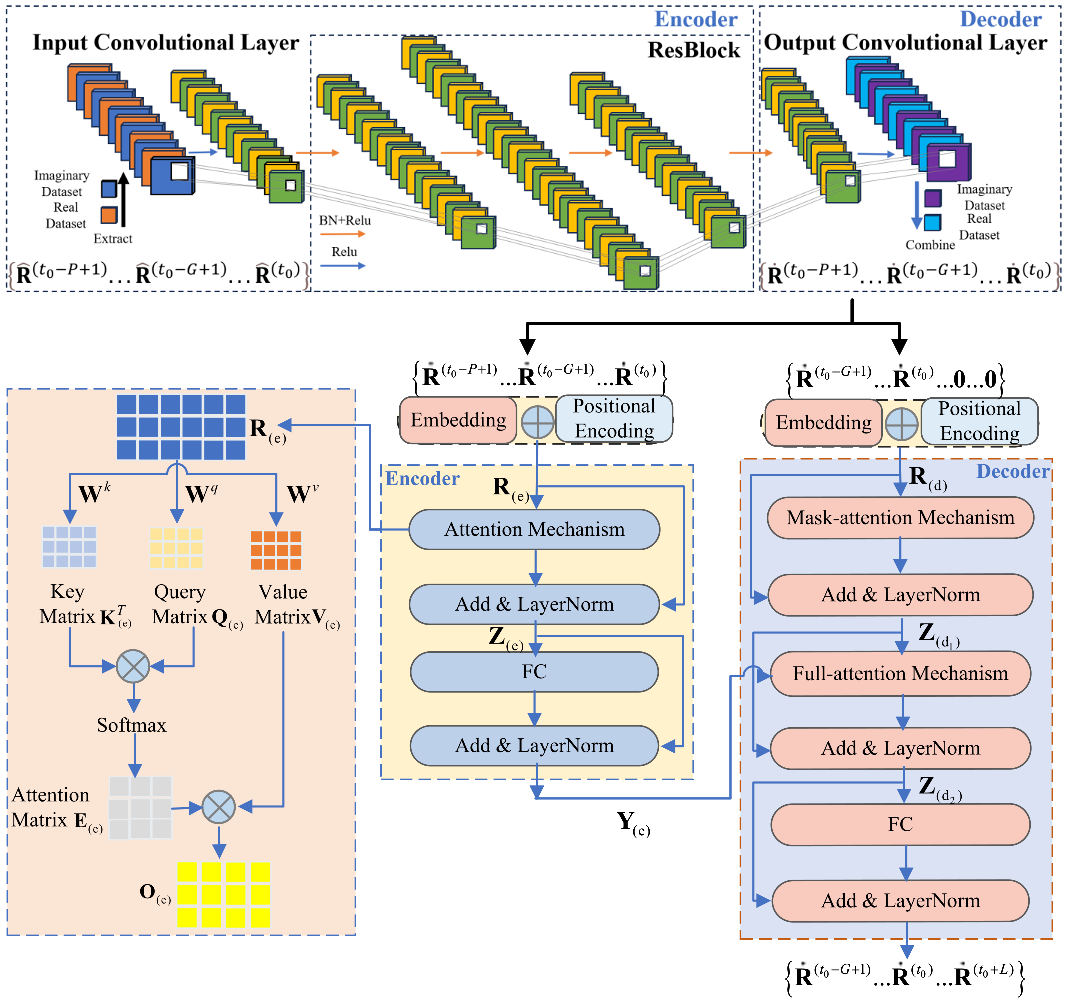}
	\caption{{The proposed 3DConvTransformer prediction model.}}\label{3DConvTransformer}
\end{figure}

After the transformation of the input IPN covariance matrices, 
$\mathbf{\tilde{R}}\in\mathbb{R}^{2K_\mathrm{A}\times K_\mathrm{A}\times X\times P}$ is fed into the 3D-CNN encoder. As shown in Fig.~\ref{3DConvTransformer}, the 3D-CNN encoder consists of an input convolutional layer and {a ResBlock}. For the input convolutional layer, $P$ filters with kernel size $K$ are leveraged to yield $P$ feature maps, and the rectified linear unit (ReLU) is utilized for regulating the gradient. {The ResBlock} consists of three convolutional layers, each with 32, 64, and $P$ filters with kernel size of $K$, respectively. The filters are followed by the batch-normalization and the ReLU. The batch-normalization is adopted to accelerate training as well as enhance denoising. The 3D-CNN decoder consists of an output convolutional layer with $P$ filters of kernel size $K$, which is used for IPN covariance matrices reconstruction. 

After the sanitization of the historically observed IPN covariance matrices, we employ a parallel transformer model to predict the IPN matrices in the future $L$ frames. As depicted in Fig.~\ref{3DConvTransformer}, the parallel transformer is composed of an embedding module, an encoder, and a decoder. The embedding module precedes the encoder and decoder stacks, which converts the IPN covariance matrices into vectors that contain positional information. Then, the encoder extracts features from historical IPN covariance vectors, capturing interrelationships and dependencies among elements of the IPN covariance vectors. By integrating encoder features with the input reference sequence, the decoder can predict IPN covariance matrices more accurately. The components of the transformer are described below in detail.

\subsubsection{Embedding} For each historical IPN covariance matrix of $\left\{\dot{\mathbf{R}}^t, t\in\left\{t_0-P+1, ..., t_0\right\}\right\}$, the real and imaginary parts are extracted, vectorized, and concatenated into a vector $\mathbf{\dot{r}}^t\in\mathbb{R}^{2K_\mathrm{A}K_\mathrm{A}X\times 1}$ as follows
\begin{equation}
\mathbf{\dot{r}}^t=[\mathrm{vec}(\mathcal{R}\{\dot{\mathbf{R}}^t\})^T,\mathrm{vec}(\mathcal{I}\{\dot{\mathbf{R}}^t\})^T]^T.
\end{equation}
After that, the sequence $\left\{\mathbf{\dot{r}}^{(t_0-P+1)}, ..., \mathbf{\dot{r}}^{t_0}\right\}$ with the size of $2K_\mathrm{A}K_\mathrm{A}X\times P$ is fed into the embedding operation module. The embedding operation consists of the original input embedding and the positional embedding, {where the original input embedding converts $\mathbf{\dot{r}}^t, \forall t\in\left\{t_0-P+1, ..., t_0\right\}$ into an $M$-dimensional vector.} The corresponding positional information is captured in the positional embedding so that the model can take advantage of the sequence order. Specifically, the position is embedded in the following equations~\cite{2017Attention} 
\begin{equation}\begin{aligned}
\text{PE}(pos,2i)&=\sin\left(\frac{pos}{10000^{\frac{2i}{M}}}\right),
\end{aligned}\end{equation}
\begin{equation}\begin{aligned}
\text{PE}(pos,2i+1)&=\cos\left(\frac{pos}{10000^{\frac{2i}{M}}}\right),
\end{aligned}\end{equation}
where $pos\in\left\{0, ..., P-1\right\}$ denotes the position of each element in the input sequence, and $i\in\left\{0, ..., \frac{M}{2}-1\right\}$ represents the current dimension of positional embedding. Consequently, $\left\{\mathbf{\dot{r}}^{(t_0-P+1)}, ..., \mathbf{\dot{r}}^{t_0}\right\}$ is processed by the embedding layer, generating an output denoted by $\mathbf{R}_{\text{(e)}}=[\mathbf{r}_{1},\cdots,\mathbf{r}_P]\in\mathbb{C}^{M \times P}$. 
\subsubsection{Encoder}
In the encoder, the input matrix $\mathbf{R}_{\text{(e)}}$, composed of IPN covariance vectors in the historical $P$ frames, is first fed into the attention layer. The attention mechanism is used to capture the global dependencies within the input IPN covariance sequence. Specifically, the trainable weight matrices $\mathbf{W}^{k}\in\mathbb{R}^{N\times M}$, $\mathbf{W}^{q}\in\mathbb{R}^{N\times M}$ and
$\mathbf{W}^{v}\in\mathbb{R}^{M\times M}$ perform linear transformations to project the raw matrix $\mathbf{R}_{\text{(e)}}$ into the key matrix
$\mathbf{K}_{\text{(e)}}\in\mathbb{R}^{N\times P}$, the query matrix $\mathbf{Q}_{\text{(e)}}\in\mathbb{R}^{N\times P}$, and the value matrix $\mathbf{V}_{\text{(e)}}\in\mathbb{R}^{M\times P}$, respectively, where $N$ represents the feature dimension of the key matrix. The key, query, and value matrices can be calculated as
\begin{equation}\label{eq:key}\begin{aligned}
\mathbf{K}_{\text{(e)}}&=\mathbf{W}^{k}\mathbf{R}_{\text{(e)}},\\\mathbf{Q}_{\text{(e)}}&=\mathbf{W}^{q}\mathbf{R}_{\text{(e)}},\\\mathbf{V}_{\text{(e)}}&=\mathbf{W}^{v}\mathbf{R}_{\text{(e)}}.
\end{aligned}\end{equation}
Subsequently, attention scores between different input IPN covariance vectors are calculated to measure the importance or correlation of each element in the historical IPN covariance vectors of the input, allowing the model to focus more on the parts that significantly impact the prediction of the IPN covariance vectors. Then, these attention scores are applied to the matrix $\mathbf{V}_{\text{(e)}}$ to obtain the final representation of the IPN covariance vectors. The attention scheme is implemented as follows
\begin{itemize}  
\item \textit{Step 1:} Compute the attention scores between different input vectors with $\mathbf{S}=\mathbf{K}_{\text{(e)}}^{T}\mathbf{Q}_{\text{(e)}}, \mathbf{S}\in\mathbb{R}^{P\times P}$; 
\item \textit{Step 2:} Normalize the attention scores for the stability of gradient with $\mathbf{S}_{\text{n}}=\frac{\mathbf{K}_{\text{(e)}}^{T}\mathbf{Q}_{\text{(e)}}}{\sqrt{N}}$;
\item \textit{Step 3:} Translate the attention scores into probabilities with
softmax function $\mathbf{E}_{\text{(e)}}=\mathrm{Softmax}(\mathbf{S}_{\text{n}})$;
\item \textit{Step 4:} Obtain the weighted value matrix with $\mathbf{O}_{\text{(e)}}=\mathbf{V}_{\text{(e)}}\mathbf{E}_{\text{(e)}}, \mathbf{O}_{\text{(e)}}\in\mathbb{R}^{M\times P}$.
\end{itemize}
The process can be unified into~\cite{10345768}
\begin{equation}
\mathbf{O}_{\text{(e)}}=\mathbf{V}_{\text{(e)}}\mathrm{Softmax}\left(\frac{\mathbf{K}_{\text{(e)}}^T\mathbf{Q}_{\text{(e)}}}{\sqrt{N}}\right).
\label{eq:OE}
\end{equation}

Then, the raw matrix $\mathbf{R}_{{\text{(e)}}}$ is added {with} the weighted value matrix $\mathbf{O}_{\text{(e)}}$ known as a residual connection, which is followed by the layer normalization (LN) operation~\cite{dai35}. The entire process is called the add and layer normalization (ADLN), which is designed to address the vanishing gradient problem~\cite{7780459}. The output of the ADLN operation is given by:
 \begin{equation}
\mathbf{Z}_{\text{(e)}}=\text{LN}\left(\mathbf{R}_{{\text{(e)}}}+\mathbf{O}_{\text{(e)}}\right).
 \end{equation}
Next, the output $\mathbf{Z}_{\text{(e)}}$ is fed into the fully connected (FC) layer. The FC layer is followed by another ADLN operation to obtain the final output $\mathbf{Y}_{\text{(e)}}$
 \begin{equation}
\mathbf{Y}_{\text{(e)}}=\text{LN}\left(\mathbf{Z}_{\text{(e)}}+\text{FC}(\mathbf{Z}_{\text{(e)}})\right).
\label{eq:YE}
 \end{equation}
\subsubsection{Decoder}
In the decoder, the input matrix $\mathbf{R}_{\text{(d)}}=[\mathbf{r}_{1},\cdots,\mathbf{r}_{G}, \mathbf{0},\cdots,\mathbf{0}]\in\mathbb{R}^{M \times (G+L)}$, composed of the IPN covariance vectors from the previous $G$ frames (where $ G\leq P$) and $L$ zero-padding vectors, is first sent into the mask-attention layer. {The masked attention} ensures that the decoder focuses on the known {values} and ignores the zero-padding ones. The key matrix $\mathbf{K}_{(\text{d}_1)}$, the query matrix $\mathbf{Q}_{(\text{d}_1)}$ and the value matrix $\mathbf{V}_{(\text{d}_1)}$ can be calculated based on the decoder's input $\mathbf{R}_{(\text{d})}$, following an approach similar to Eq.~\eqref{eq:key}. {On the basis of the attention scores obtained from the dot product of $\mathbf{K}_{(\text{d}_1)}^{T}$ and $\mathbf{Q}_{(\text{d}_1)}$,} a mask matrix $\mathbf{C}\in\mathbb{R}^{(G + L)\times (G + L)}$ is applied to ensure that the model only focuses on the past and current IPN covariance vectors. The matrix $\mathbf{C}$ masks the last $L$ frames from the input layer by setting the last 
$k$ rows of $\mathbf{C}$ to $-\infty$, which can be defined as
\begin{equation}
\mathbf{C}_{i,j}=\begin{cases}-\infty&\text{if}\quad i \ge {G + L}-k+1,\\1&\text{otherwise}.\end{cases}
\end{equation}
Subsequently, the obtained attention scores are used to repeat \textit{Steps 2-4} in the attention mechanism, thereby yielding the output of the mask attention. With the ADLN operation, we obtain 
\begin{equation}
\mathbf{Z}_{(\text{d}_1)}=\text{LN}\left(\mathbf{R}_{(\text{d})}+\mathbf{V}_{(\text{d}_1)}\text{Softmax}\left(\frac{(\mathbf{K}_{(\text{d}_1)})^{T}\mathbf{Q}_{(\text{d}_1)}}{\sqrt{N}}\mathbf{C}\right)\right).
\label{eq:ZD1}
\end{equation}
The subsequent full-attention layer combines $\mathbf{Z}_{(\text{d}_1)}$ with the encoder's feature matrix $\mathbf{Y}_{\text{(e)}}$, which further enables the decoder to focus on the critical components of the input sequence when generating each IPN covariance vector. The key matrix 
$\mathbf{K}_{(\text{d}_2)}$, the value matrix 
$\mathbf{V}_{(\text{d}_2)}$, and the query matrix $\mathbf{Q}_{(\text{d}_2)}$ are calculated as
\begin{equation}
\label{eq:keyd2}
\begin{aligned}
\mathbf{K}_{(\text{d}_2)}&=\mathbf{W}^{k}_2\mathbf{Y}_{\text{(e)}},\\\mathbf{Q}_{(\text{d}_2)}&=\mathbf{W}^{q}_2\mathbf{Z}_{(\text{d}_1)},\\\mathbf{V}_{(\text{d}_2)}&=\mathbf{W}^{v}_2\mathbf{Y}_{\text{(e)}},
\end{aligned}
\end{equation}
where $\mathbf{W}_2^{k}\in\mathbb{R}^{N\times M}$, $\mathbf{W}_2^{q}\in\mathbb{R}^{N\times M}$ and
$\mathbf{W}_2^{v}\in\mathbb{R}^{M\times M}$ are the trainable weight matrices in the full-attention layer. With another ADLN operation, we obtain
\begin{equation}
\mathbf{Z}_{(\text{d}_2)}=\text{LN}\left(\mathbf{Z}_{(\text{d}_1)}+\mathbf{V}_{(\text{d}_2)}\text{Softmax}\left(\frac{(\mathbf{K}_{(\text{d}_2)})^{T}\mathbf{Q}_{(\text{d}_2)}}{\sqrt{N}}\right)\right).
\label{eq:ZD}
\end{equation}
The next layer is the FC layer, followed by another ADLN operation, producing the final output
\begin{equation}
\mathbf{Y}_{\text{(d)}}=\mathrm{LN}\left(\mathbf{Z}_{(\text{d}_2)}+\text{FC}(\mathbf{Z}_{(\text{d}_2)})\right).
\label{eq:YD}
\end{equation}
Then, we reshape $\mathbf{Y}_{\text{(d)}}$ into $\dot{\mathbf{R}}^{t}\in\mathbb{R}^{(G+L)\times X\times K_{\mathrm{A}}\times K_{\mathrm{A}}}$ by combining the corresponding real and imaginary parts. By extracting the last $L$ columns of $\dot{\mathbf{R}}^{t}$, we obtain the final predicted IPN covariance matrix $\left\{\dot{\mathbf R}^{(t_0+1)},...,\dot{\mathbf{R}}^{(t_0+L)}\right\}$.

Compared to MSE, normalized mean square error (NMSE) is a more suitable metric to interpret and compare across different datasets with the normalization of the error. Additionally, the normalized loss can yield a steady gradient, and thus accelerate the convergence. Therefore, we utilize the NMSE between the predicted IPN covariance matrices $\left\{{\dot{\mathbf{{R}}}^{(t_0+1)}},...,{\dot{\mathbf{{R}}}^{(t_0+L)}}\right\}$ and the accurate ones $\left\{\mathbf{{R}}^{(t_0+1)},...,\mathbf{{R}}^{(t_0+L)}\right\}$ as the loss function of the proposed 3DConvTransformer model, which can be expressed as
\begin{equation}
\mathcal{L}_{\text{3DConvTran}}=\mathbb{E}\left\{\frac{\sum_{t=t_0+1}^{t_0+L}\|{\mathbf{R}}^t-\dot{\mathbf{R}}^t\|^{2}}{\sum_{t=t_0+1}^{t_0+L}\|{\mathbf{R}}^t\|^{2}}\right\},
\label{eq:3Dloss-function}
\end{equation}
where the output $\left\{\dot{{\mathbf{{R}}}}^{(t_0+1)},...,\dot{{\mathbf{{R}}}}^{(t_0+L)}\right\}$ is the mapping from the 3DConvTransformer input and the training parameters set ${\Theta} = \left\{\Theta_{\text{conv}},\Theta_{\text{trans}}\right\}$ with $\Theta_{\text{trans}}$ representing the training parameters set of the transformer. The mapping function is denoted by $\mathcal{F}$ as follows 
\begin{equation}
\left\{\dot{{\mathbf{{R}}}}^{(t_0+1)},...,\dot{{\mathbf{{R}}}}^{(t_0+L)}\right\} = \mathcal{F}\left(\left\{\dot{{\mathbf{{R}}}}^{(t_0-P+1)},...,\dot{{\mathbf{{R}}}}^{t_0}\right\}, {\Theta}\right).
\end{equation}  

The procedure of the 3DConvTransformer is summarized in Algorithm~\ref{alg:3DConvTransformer}, where $\hat{\mathbf{R}}$ is firstly reshaped into a real-valued matrix $\tilde{\mathbf{R}}$ in line 1. Next, the IPN covariance matrices from the past $P$ frames are processed with the 3D-CNN for feature extraction in line 2. Subsequently, the denoised IPN covariance matrices are fed into the Transformer encoder and decoder for further processing in lines {3-9}. Then, the output is reshaped to the complex-valued matrix in line 10 and the IPN covariance matrices of the future $ L$ frames are extracted as the final output. 

\begin{algorithm}[!t]
\caption{The proposed 3DConvTransformer algorithm}
\label{alg:3DConvTransformer}
\LinesNumbered

\KwIn{IPN covariance matrices in the previous $P$ frames 
$\left\{\mathbf {\hat{R}}^{(t_0-P+1)},..., \mathbf {\hat{R}}^{t_0}\right\}$.} 
\KwOut{Predicted IPN covariance matrices in the future $L$ frames $\left\{\dot{{\mathbf{{R}}}}^{(t_0+1)},...,\dot{{\mathbf{{R}}}}^{(t_0+L)}\right\}$.

}

        Reshape $\hat{\mathbf{R}}$ into a real-valued matrix $\tilde{\mathbf{R}}$ by extracting the real and imaginary parts\;
        Extract the features of $\tilde{\mathbf{R}}$ with the 3D-CNN\;
        \textbf{Transformer Encoder:}
        
         Calculate $\mathbf{O}_{(\text{e})}$ by the attention layer based on Eq.~\eqref{eq:OE}\;   
         Obtain $\mathbf{Y}_{(\text{e})}$ by the FC layer and the ADLN based on Eq.~\eqref{eq:YE}\;         \textbf{Transformer Decoder:}
         
         Derive $\mathbf{Z}_{(\text{d}_1)}$ by the mask-attention layer and the ADLN based on Eq.~\eqref{eq:ZD1}\;
        Generate $\mathbf{Z}_{(\text{d}_2)}$ by the full-attention layer and the ADLN based on Eq.~\eqref{eq:ZD}\;
        Compute $\mathbf{Y}_{\text{(d)}}$ by the FC layer and the ADLN based on Eq.~\eqref{eq:YD}\;
        Reshape $\mathbf{Y}_{\text{(d)}}$ into a complex-valued matrix $\dot{\mathbf{R}}^t$ by combining the real and imaginary parts\;
Output $\left\{\dot{{\mathbf{{R}}}}^{(t_0+1)},...,\dot{{\mathbf{{R}}}}^{(t_0+L)}\right\}$ by extracting the last $L$ columns of $\dot{\mathbf{R}}^t$.\
\end{algorithm}
\subsection{Complexity Analysis} Next, we analyze the computational complexity in terms of the number of matrix multiplications in Algorithm~\ref{alg:3DConvTransformer}. The complexity of the 3D-CNN is dominated by the convolutional module, which consists of an input convolutional layer, a ResBlock with three convolutional layers and an output convolutional layer. Accordingly, the complexity is given as 
$\mathcal{O}\left(KK_\mathrm{A}^2X \left(P^2+\sum_{a=1}^A 
C_{a}^{\text{in}}C_{a}^{\text{out}}\right)\right)$,
where $A$ denotes the total number of convolutional layers in ResBlock, $C_{a}^{\text{in}}$ and $C_{a}^{\text{out}}$ denote the number of input and output channels of the $a$-th layer, respectively. The complexity of the transformer-based parallel prediction model is dominated by the attention mechanism and the FC processing. On the one hand, the complexity of the attention mechanism is determined by the matrix multiplications involved in key, query, and value projections, which is given by $\mathcal{O}\left(K_\mathrm{A}^4X^2+(G+P+L)K_\mathrm{A}^2X\right)$~\cite{9832933}. On the other hand, the complexity of the FC processing is given by $\mathcal{O}\left(K_\mathrm{A}^4X^2\right)$. Therefore, the overall complexity of the 3DConvTransformer is given as $\mathcal{O}\left(KK_\mathrm{A}^2X \left(P^2+\sum_{a=1}^A 
C_{a}^{\text{in}}C_{a}^{\text{out}}\right)\right)+\mathcal{O}\left(K_\mathrm{A}^4X^2+\left(G+P+L\right)K_\mathrm{A}^2X\right)$.

\section{Interference-Resistant Hybrid Beamforming}
In this section, we focus on the IR-HBF design given the predicted IPN covariance matrices. Since the optimization problem \eqref{eq:optimization_problem} involves the joint optimization over five coupled variables, i.e., baseband precoder $\mathbf{V}_{\text{BB},x}$, RF precoder $\mathbf{V}_{\text{RF,}}$, RF combiner $\mathbf{W}_{\text{RF}}$, baseband combiner $\mathbf{W}_{\text{BB},x}$, and scaling
factor ${\beta_{x}}$, along with non-convex constraints, it is intractable to find the optimal solution. In the following, we first introduce the conventional iteration-based AO approach, and then elaborate on the novel KDDD IR-HBF in details. 

\subsection{AO Approach}
To address the MSE minimization problem given in~\eqref{eq:optimization_problem}, the well-known approach is the AO where $\mathbf{V}_{\text{BB},x}$, $\mathbf{V}_{\text{RF,}}$, $\mathbf{W}_{\text{RF}}$, $\mathbf{W}_{\text{BB},x}$ are decoupled and solved iteratively~\cite{8616797}. The iterations can be processed hierarchically via the outer and inner loops, with the outer loop iterating between the precoder and the decoder, and the inner loop iterating between the baseband and the RF processing. Compared to the formulated problem in~\cite{8616797}, the main difference of our work is the existence of interference. Therefore, we focus on the explanation of the decoder design for the spatial-domain interference suppression, while the precoder design subproblem can be resolved in an easier approach. Specifically, by fixing the precoding matrix $\mathbf{V}_{x}$ in Eq.~\eqref{eq:MSE}, the combiner design subproblem can be formulated as 
\begin{subequations}
\label{eq:combiner-optimization-problem}
    \begin{align}
        \min _{\mathbf{W}_{\text{RF}},\mathbf{W}_{\text{BB},x}}& \sum_{x=0}^{X-1}\text{Tr} \left(\mathbf{W}_{\text{BB},x}^H\mathbf{W}_\text{RF}^H\mathbf{H}_1\mathbf{H}_1^H\mathbf{W}_\text{RF}\mathbf{W}_{\text{BB},x}\right.\nonumber\\  
        & -\mathbf{W}_{\text{BB},x}^H\mathbf{W}_\text{RF}^H\mathbf{H}_1-\mathbf{H}_1^H\mathbf{W}_\text{RF}\mathbf{W}_{\text{BB},x}\nonumber\\  
        &+        \left.\beta^{-2}_x\mathbf{W}_{\text{BB},x}^H\mathbf{W}_\text{RF}^H\mathbf{R}_x\mathbf{W}_\text{RF}\mathbf{W}_{\text{BB},x}+\mathbf{I}_{N_s}\right),
    \end{align}
    \begin{equation}
    \label{eq:combiner_RF_constraint}
        {\rm{s.t.}} \quad \quad
|[\mathbf{W_{\text{RF}}}]_{m,l}|=1,\forall m,l,
    \end{equation}
\end{subequations}
where $\mathbf{H}_1 \triangleq \beta^{-1}_x{\mathbf{H}}_{\text{BA}}^x\mathbf{V}_{x}$. Observing \eqref{eq:combiner-optimization-problem}, one can find that the objective function is convex with respect to the independent baseband combiners $\mathbf{W}_{\text{BB},x}, \forall x\in\left\{1, ..., X\right\}$. With some simple differential operations, the optimal baseband combiner over the $x$-th subcarrier can be obtained as
\begin{equation}
    \label{eq:combiner}
    \mathbf{W}_{\text{BB},x}=\left(\mathbf{W}_{\text{RF}}^{H}\widetilde{\mathbf{H}}_1\mathbf{W}_{\text{RF}}+\beta_x^{-2}\mathbf{W}_{\text{RF}}^{H}\mathbf{R}_x
        \mathbf{W}_{\text{RF}}\right)^{-1}
        \mathbf{W}_{\text{RF}}^{H}\mathbf{H}_{1},
\end{equation}
where $\widetilde{\mathbf{H}}_1 \triangleq \mathbf{H}_{1}\mathbf{H}_{1}^{H}$. With the optimal $\mathbf{W}_{\text{BB},x}$, the resulting MSE can be rewritten as
\begin{align}
    J_x(\mathbf{W}_{\text{RF}}) = \mathrm{Tr}\left(\mathbf{I}_{N_{\text{s}}}+\beta^2_x\mathbf{H}_1^H\mathbf{W}_{\text{RF}}\mathbf{\Gamma}_x^{-1}\mathbf{W}_{\text{RF}}^H\mathbf{H}_1\right)^{-1},
\end{align}
where $\mathbf{\Gamma}_x \triangleq \mathbf{W}_{\text{RF}}^H\mathbf{R}_x\mathbf{W}_{\text{RF}}$.
Correspondingly, the original problem ~\eqref{eq:combiner-optimization-problem} is simplified and transferred to the following one for the optimization of $\mathbf{W}_{\text{RF}}$
 \begin{subequations}
 \label{eq:optimal-RF}
 \begin{align}
 \underset{\mathbf{W_{\text{RF}}}}{\operatorname*{min}}\quad &\sum_{x=0}^{X-1}J_x(\mathbf{W_{\text{RF}}})\\
 {\rm{s.t.}}\quad &
 |[\mathbf{W_{\text{RF}}}]_{m,l}|=1,\forall m,l.\end{align}
 \end{subequations}
In contrast to the closed-form derivation of $\mathbf{W}_{\text{BB},x}$, \eqref{eq:optimal-RF} is a non-convex problem due to the constant-modulus constraint. The optimization of~$\mathbf{W}_{\text{RF}}$ should be executed upon the complex-circle manifold $\mathcal{W} = \left\{\mathbf{W_{\text{RF}}}: \left|\left[\mathbf{W_{\text{RF}}}\right]_{m,l}\right|=1\right\}$, which is a typical MO problem~\cite{manifold-optimization1,manifold-optimization2}. To be more specific, with the initialized point located at the manifold, $\mathbf{W}_{\text{RF}}$ is iteratively updated in the direction of the Riemannian gradient until approximating the optimal solution. To obtain the Riemannian gradient, we first derive the conjugate gradient in the Euclidean space as follows
\begin{equation}\begin{aligned}
\nabla J_x(\mathbf{W}_{\text{RF}})& =\frac{\partial J_x(\mathbf{W}_\text{RF})}{\partial\mathbf{W_\text{RF}^{*}}}  \\
&=\beta^2_x\left(\mathbf{R}_x\mathbf{W}_{\text{RF}}\mathbf{\Gamma}_x^{-1}\mathbf{W}_{\text{RF}}^H\mathbf{H}_1\mathbf{G}_x^{-2}\mathbf{H}_{1}^H\mathbf{W}_{\text{RF}}\mathbf{\Gamma}_x^{-1} \right.\\
&-\left.\mathbf{H}_1\mathbf{G}_x^{-2}\mathbf{H}_{1}^H\mathbf{W}_{\text{RF}}\mathbf{\Gamma}_x^{-1}\right),
\end{aligned}\end{equation}
where $\mathbf{G}_x\triangleq \mathbf{I}_{N_{\text{s}}}+\beta^2_x\mathbf{H}_1^H\mathbf{W}_{\text{RF}}\mathbf{\Gamma}_x^{-1}\mathbf{W}_{\text{RF}}^H\mathbf{H}_1$.
With the derived Euclidean conjugate gradient, we have the following steps: 
\begin{itemize}
\item \textit{Step 1}: Project the Euclidean conjugate gradient onto the tangent space to obtain the Riemannian gradient; 
\item \textit{Step 2}: Apply the Armijo backtracking line search~\cite{manifold-optimization2} to determine the step size, so as to update $\mathbf{W}_{\text{RF}}$ in the tangent space; 
\item \textit{Step 3}: Retract  $\mathbf{W}_{\text{RF}}$ back to the manifold; 
\item \textit{Step 4}: Repeat \textit{Steps 1-3} until convergence. 
\end{itemize}

For the precoder design, the above closed-form derivation of the baseband processing and the MO-based analog processing can be directly applied, only except that the scaling parameter $\beta_x$ needs to be jointly optimized with the precoder considering the IPN effect~\cite{linear-transmit}. Fortunately, the MSE function is convex with respect to $\beta_x$, and therefore the closed-form expression can be derived. For more details, the reader is referred to~\cite{8616797}.

\subsection{KDDD Approach}
The conventional AO approach is time consuming due to the existence of the nested loop. In the rapidly-varying communications scenarios, the computation time is restricted by the maximum tolerable latency. As such, our aim is to boost the performance within the limited number of iterations. One can observe that the performance of the AO highly depends on the choice of the gradient descent step sizes in the MO, which inspires us to treat the step sizes as trainable parameters. To be more specific, by getting access to the simulated channel and IPN realizations, the step sizes are learned intelligently to minimize the loss function, i.e., the MSE defined in~\eqref{eq:MSE}. Compared to the Armijo backtracking line search, the learning-based approach can find a generalized set of step sizes to approximate the optimal performance under the predefined number of outer and inner iterations. Moreover, a common drawback of the gradient descent algorithm is that it can easily get stuck in the local optimum. This problem can be avoided by deploying the ``empiricism" obtained via a large number of training samples. Inspired by the above discussion, we propose to construct a targeted network, i.e., the KDDD-IRN network, where the structure is visibly simplified compared to the conventional ``black-box" NN, by replicating the AO operations.

The proposed KDDD-IRN structure is shown in Fig.~\ref{a}. We map the outer iterations of the AO to $I$ KDDD-IRN layers, where each layer contains one iteration between the update of the precoder and the combiner. There are four blocks, i.e., the RF precoder, the BB precoder, the RF combiner and the BB combiner, within each layer. For the RF precoder/combiner blocks, the Riemannian gradient descent is executed for $J$ steps by incorporating the step sizes as trainable parameters. Specifically, the operations of each Riemannian gradient descent step in the RF precoder and combiner are shown in Fig.~\ref{b} and Fig.~\ref{c}, respectively. By replicating the operations of the typical AO algorithm, it is clear that the KDDD-IRN has much simpler structure and less trainable parameters compared to its ``black-box" counterpart. Specifically, we denote the BB precoder, the RF precoder, the BB combiner, and the RF combiner after the $j$-th gradient descent step in the $i$-th layer by $\mathbf{V}_{\text{BB},x}^{i,j}$, $\mathbf{V}_{\text{RF}}^{i,j}$, 
$\mathbf{W}_{\text{BB},x}^{i,j}$, $\mathbf{W}_{\text{RF}}^{i,j}$, respectively, and the set of trainable parameters by $\left\{\gamma_{\text{B}}^{i},\gamma_{\text{A}}^{i}, \forall i\in\left\{1, ..., I\right\}\right\}$, with
$\gamma_{\text{B}}^{i} = \left\{\gamma_{\text{B}}^{i,1},\gamma_{\text{B}}^{i,2},...\gamma_{\text{B}}^{i,J}\right\}$ and $\gamma_{\text{A}}^{i}=\left\{\gamma_{\text{A}}^{i,1},\gamma_{\text{A}}^{i,2},...\gamma_{\text{A}}^{i,J}\right\}$ representing the set of step sizes over the $J$ gradient descent steps in the $i$-th layer. We further describe the four blocks of the $i$-th layer with details in the following. 

\textit{1) RF precoder block}: The block contains $J$ gradient descent steps with the input $\mathbf{V}_{\text{RF}}^{i-1,J}$ as the initial setting. The RF precoder is updated upon the tangent space with the $j$-th Riemannian gradient  $\nabla_{\mathrm{R}}\left(J_x(\mathbf{V}_{\text{RF}}^{i,j})\right)$ as 
\begin{equation}
\label{eq:gradient}
\overline{\mathbf{V}}_{\text{RF}}^{i,j+1}=\mathbf{V}_{\text{RF}}^{i,j}-\gamma_{\text{B}}^{i,j}\frac{\nabla_{\mathrm{R}}\left(J_x(\mathbf{V}_{\text{RF}}^{i,j})\right)}{\|\nabla_{\mathrm{R}}\left(J_x(\mathbf{V}_{\text{RF}}^{i,j})\right)\|_{F}}.
\end{equation}
Afterwards, $\overline{\mathbf{V}}_{\text{RF}}^{i,j+1}$ is retracted onto the complex-circle manifold as 
\begin{equation}
\begin{aligned}
\mathbf{V}_{\text{RF}}^{i,j+1}(m,n) 
&=\mathrm{Retr}\left(\overline{\mathbf{V}}_{\text{RF}}^{i,j+1}\left(m,n\right)\right)  \\
&=\frac{\overline{\mathbf{V}}_{\text{RF}}^{i,j+1}\left(m,n\right)}{|\overline{\mathbf{V}}_{\text{RF}}^{i,j+1}\left(m,n\right)|}.
\end{aligned}
\end{equation}
The output after the $J$-th iteration is $\mathbf{V}_{\text{RF}}^{i,J}$.

\textit{2) BB precoder block}: The block directly applies the closed-form solution as shown in Eq.~\eqref{eq:V_BB} for obtaining $\mathbf{V}_{\text{BB},x}^{i,J}, \forall x\in\left\{1, ..., X\right\}$, for which the detailed derivation can be found in~\cite{8616797}. The superscript $J$ indicates that the BB precoding matrix is calculated based on $\mathbf{V}_{\text{RF}}^{i,J}$. 
\begin{equation}
\label{eq:V_BB}
\begin{aligned}
&\mathbf{V}_{\text{BB},x}^{i,J}  
=\beta_x^{i,J}\widetilde{\mathbf{V}}_{\text{BB},x}^{i,J},\\
&\beta_x^{i,J}=\left(\text{Tr}\left(\mathbf{V}_{\text{RF}}^{i,J}\widetilde{\mathbf{V}}_{\text{BB},x}^{i,J}\left(\widetilde{\mathbf{V}}_{\text{BB},x}^{i,J}\right)^H\left(\mathbf{V}_{\text{RF}}^{i,J}\right)^H\right)\right)^{-\frac{1}{2}}, \\  
   &\widetilde{\mathbf{V}}_{\text{BB},x}^{i,J} = \left(\mathbf{H}_2\mathbf{H}_2^H+ \eta_x\left(\mathbf{V}_{\text{RF}}^{i,J}\right)^H\mathbf{V}_{\text{RF}}^{i,J}\right)^{-1}\mathbf{H}_2,\\ &\mathbf{H}_2\triangleq \left(\mathbf{V}_{\text{RF}}^{i,J}\right)^H\left({\mathbf{H}}_{\text{BA}}^x\right)^H\mathbf{W}_x^{i-1,J},\\   &\eta_x\triangleq\text{Tr}\left(\left(\mathbf{W}_x^{i-1,J}\right)^H\mathbf{R}_x\mathbf{W}_x^{i-1,J}\right).
\end{aligned} 
\end{equation}

\textit{3) RF combiner block}: The RF combiner block is similar to the RF precoder block by updating $\mathbf{W}_{\text{RF}}^{i,j+1}$ with $J$ gradient descent steps upon the complex-circle manifold. The output is $\mathbf{W}_{\text{RF}}^{i,J}$.

\textit{4) BB combiner block}: {This block executes the closed-form derivation for obtaining $\mathbf{W}_{\text{BB}, x}^{i,J}, \forall x\in\left\{1, ..., X\right\}$ based on Eq.~\eqref{eq:combiner}. The superscript $J$ indicates that the BB combining matrix is calculated based on $\mathbf{W}_{\text{RF}}^{i,J}$.
\begin{figure}[H]
\centering
\subfigure[]{\label{a}
\includegraphics[scale=0.35]{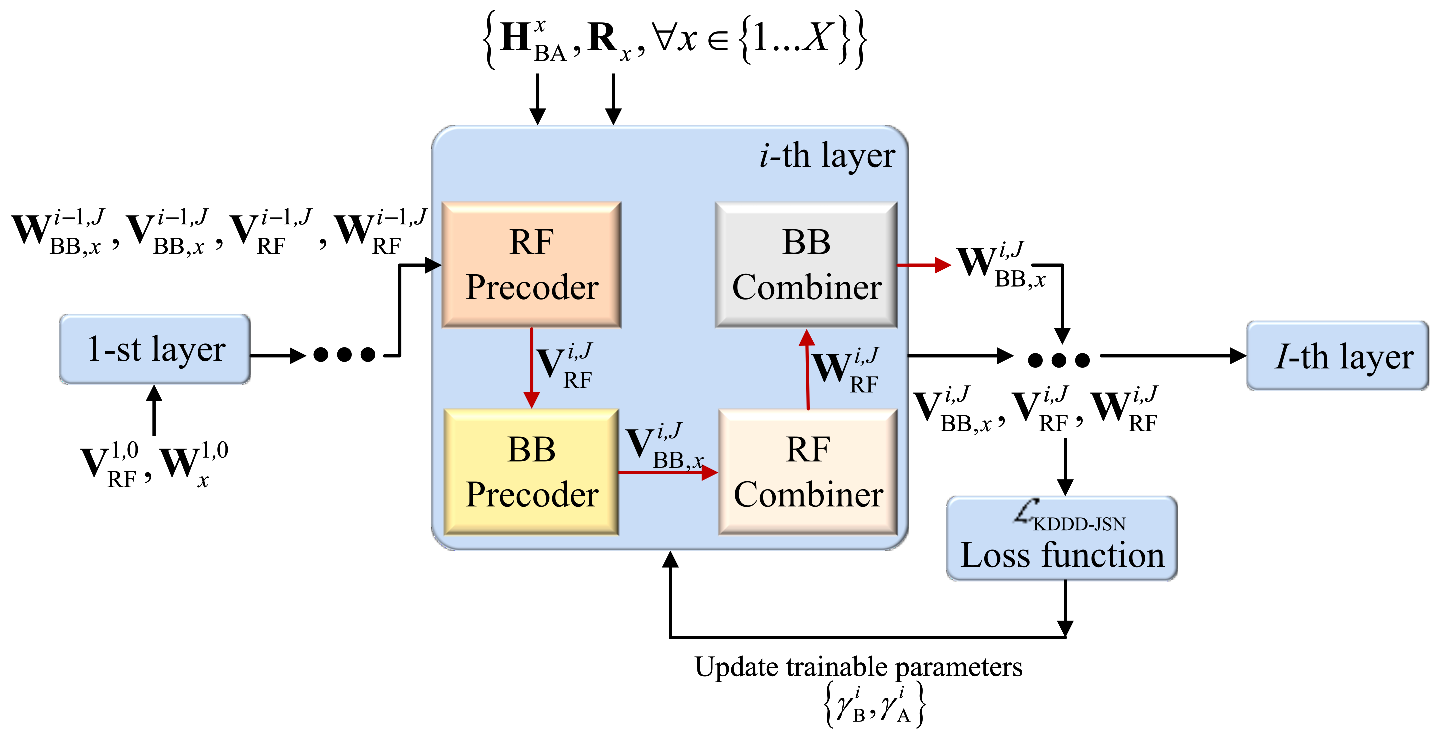}}
\subfigure[]{\label{b}
\includegraphics[scale=0.5]{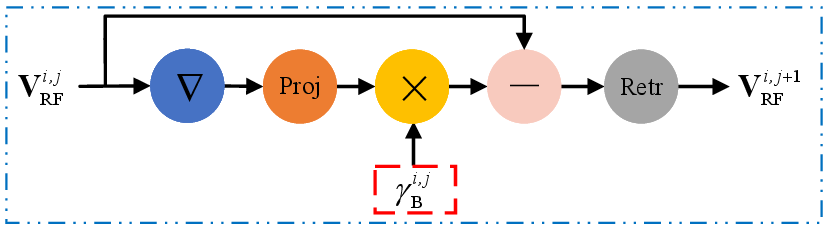}}
\subfigure[]{\label{c}
\includegraphics[scale=0.5]{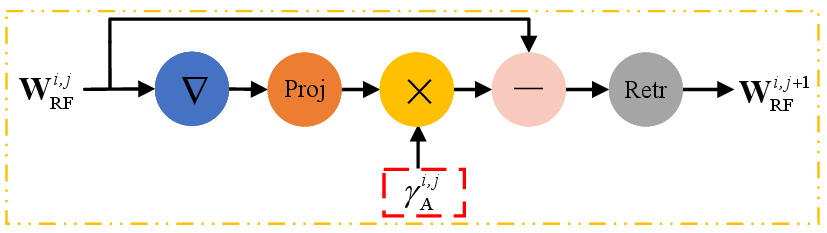}}
\caption{(a) KDDD-IRN architecture; (b) the $j$-th gradient descent step of the RF precoder; (c) the $j$-th gradient descent step of the RF combiner.}
\label{DU-HBF}
\end{figure}
The procedure of the KDDD-IRN is summarized in Algorithm~\ref{alg:DU-HBF}, which follows the unsupervised training approach. Specifically, the channel matrix ${\mathbf{H}}_{\text{BA}}^x$ and the IPN covariance matrix ${\mathbf{R}}_x$ obtained from Section III are taken as the input, which are indispensable for the calculation of the Riemannian gradient as shown in Eq.~\eqref{eq:gradient} and the loss function given as
\begin{equation}
\mathcal{L}_{\text{KDDD-IRN}} = \text{MSE}\left(\boldsymbol{\gamma}_{\text{B}}^{i},\boldsymbol{\gamma}_{\text{A}}^{i}, \forall i\in\left\{1, ..., I\right\}\right),
    \label{eq:loss-function}
\end{equation}
where $\left\{\boldsymbol{\gamma}_{\text{B}}^{i},\boldsymbol{\gamma}_{\text{A}}^{i}, \forall i\in\left\{1, ..., I\right\}\right\}$ is the set of trainable parameters, with $\boldsymbol{\gamma}_{\text{B}}^{i}=\left\{\gamma_{\text{B}}^{i,1}, ..., \gamma_{\text{B}}^{i,J}\right\}$ and $\boldsymbol{\gamma}_{\text{A}}^{i}=\left\{\gamma_{\text{A}}^{i,1}, ..., \gamma_{\text{A}}^{i,J}\right\}$ denote the set of step sizes in the $i$-th layer for the precoder and combiner update, respectively. Moreover, as the iterations among the four HBF matrices require proper initialization, we propose to employ the fully-digital interference resistant (FD-IR) beamforming for the initialization in the first layer, as the optimal HBF will finally approach the fully-digital one. To improve the performance of the KDDD-IRN, we adopt a layer-by-layer training strategy. For the layer $i>1$, the output from the $(i-1)$-th layer is taken as the input, as shown in lines {7-9}. The RF/BB precoder/combiner in the $i$-th layer are updated in lines {10-15}. Based on the loss function defined in Eq.~\eqref{eq:loss-function}, the step sizes are updated by deploying the back propagation mechanism in lines {16-17}. After completing the training process, the trained parameters are reserved as shown in line 19. In the online operation stage, $\mathbf{V}_{\text{BB},x}^{I,J}$, $\mathbf{V}_{\text{RF,}}^{I,J}$, $\mathbf{W}_{\text{RF}}^{I,J}$, $\mathbf{W}_{\text{BB},x}^{I,J}$ can be obtained directly with the predefined mapping as shown in line 20, thereby improving the online operating efficiency.}

\begin{algorithm}[h]
\caption{The proposed KDDD-IRN algorithm}
\label{alg:DU-HBF}
\LinesNumbered
\KwIn{ ${\mathbf{H}}_{\text{BA}}^x$, ${\mathbf{R}}_x$, $\mathbf{V}_{\text{RF}}^{1,0}$,  $\mathbf{W}_{x}^{1,0}, \forall x$. }
\KwOut{ ${\mathbf{V}}_{\text{BB},x}^{I,J}, {\mathbf{W}}_{\text{BB},x}^{I,J}, \mathbf{V}_{\text{RF}}^{I,J}, \mathbf{W}_{\text{RF}}^{I,J}, \forall x$.}

\For{i = $1, ..., I$}{
        Initialize the trainable parameters $\left\{\boldsymbol{\gamma}_{\text{B}}^{i},\boldsymbol{\gamma}_{\text{A}}^{i}, \forall i\right\}$\;
        {Initialize training epochs $E$}\;
        Initialize the batch size $B$\;
        \For{e = $1, ..., E$}{
            \For{b = $1, ..., B$}{
             \If {i \textgreater $1$}{
        $\mathbf{V}_{\text{RF}}^{i,0} = \mathbf{V}_{\text{RF}}^{i-1,J}, 
        \mathbf{V}_{\text{BB},x}^{i,0} = \mathbf{V}_{\text{BB},x}^{i-1,J}$\;
        $\mathbf{W}_{\text{RF}}^{i,0} = \mathbf{W}_{\text{RF}}^{i-1,J}, 
        \mathbf{W}_{\text{BB},x}^{i,0} = \mathbf{W}_{\text{BB},x}^{i-1,J}$\;
    }
             \For{j = $1, ..., J$}{
                Compute the RF precoder matrix $\mathbf{V}_{\text{RF}}^{i,j}$ with $\gamma_{\text{B}}^{i,j}$;}
                Compute the BB precoder matrix ${\mathbf{V}}_{\text{BB},x}^{i,J}, \forall x$\;
            \For{j = $1, ..., J$}{
                Compute the RF combiner matrix $\mathbf{W}_{\text{RF}}^{i,j}$ with $\gamma_{\text{A}}^{i,j}$;
            }
                Compute the BB combiner matrix ${\mathbf{W}}_{\text{BB},x}^{i,J}, \forall x$\;
     }
     Compute the average loss of the batch based on Eq.~\eqref{eq:loss-function}\;
                Update $\{\boldsymbol{\gamma}_{\text{B}}^{i},\boldsymbol{\gamma}_{\text{A}}^{i}\}$ with the back propagation mechanism;
           
        }
        \textbf{return} trained $\{\boldsymbol{\gamma}_{\text{B}}^{i},\boldsymbol{\gamma}_{\text{A}}^{i}\}$;
}
Reserve trained parameter $\left\{\boldsymbol{\gamma}_{\text{B}}^{i}, \boldsymbol{\gamma}_{\text{A}}^{i}, \forall i\right\}$;

\textbf{return} $\mathbf{V}_{\text{RF}}^{I,J}, {\mathbf{V}}_{\text{BB},x}^{I,J}, \mathbf{W}_{\text{RF}}^{I,J},{\mathbf{W}}_{\text{BB},x}^{I,J}, \forall x$ for online operation.

\end{algorithm}

Next, we analyze the computational complexity of the proposed KDDD-IRN. Since the precoder and combiner designs follow a similar approach, we focus on the complexity analysis of the combining matrices calculation. Moreover, since the digital processing is with simpler solving procedure as well as smaller matrix manipulation dimensions compared to the analog counterpart, the computational complexity analysis of the digital processing is neglected here.
As shown in Fig.~\ref{a}, the main parts for complicated matrices operations in the RF combiner block include the calculation of the conjugate gradient, the projection of the conjugate gradient to the tangent space, as well as the retraction operation. For simplification, we denote $K_{\text{RF}}=K_{\text{RF}}^{\text{A}}=K_{\text{RF}}^{\text{B}}$. Then the complexity of the conjugate gradient calculation is given by $X\left(N_{\text{s}}^2K_{\text{B}}+2K_{\text{RF}}^2K_{\text{A}}+K_{\text{RF}}K_{\text{A}}^2+6N_{\text{s}}K_{\text{A}}K_{\text{RF}}+2N_{\text{s}}K_{\text{RF}}^2+\right.\\
\left.3N_{\text{s}}^2K_{\text{A}}+N_{\text{s}}^3+\mathcal{O}\left(K_{\text{RF}}^3\right)+\mathcal{O}\left(N_{\text{s}}^3\right)\right)$, where
$\mathcal{O}\left(K_{\text{RF}}^3\right)+\mathcal{O}\left(N_{\text{s}}^3\right)$ results from the matrix inversions, while the remaining items result from the matrix multiplications. 
The orthogonal projection is essentially the Hadamard product which takes $X\left(2K_{\text{A}}K_{\text{RF}}\right)$ multiplications.
In addition, the complexity of the retraction
operation is $X\left(K_{\text{A}}K_{\text{RF}}\right)$. 
\begin{remark}
    It is worth noting that, with the same number of iterations, the complexity of the KDDD-IRN is lower than that of the conventional AO algorithm,
    relying on the fact that the data-fed learning is adopted to substitute the complicated backtracking line search for the step sizes determination. We will provide the quantitative complexity comparison between the KDDD-IRN and the AO in Section~V.
\end{remark}

\section{Numerical Results}
In this section, we evaluate the performance of the proposed interference-robust MIMO communications scheme, in comparison with state-of-the-art techniques. 

\subsection{Simulation Setup}
Unless otherwise stated, we set the number of subcarriers as $X=64$, the number of transmitted data streams as $N_\text{s} = 2$. The AC and LGS are both assumed to be equipped with a UPA of size $16$, i.e., $K_\text{A} = K_\text{B}=16$, and $4$ RF chains, i.e., $K_{\text{RF}}^{\text{A}}=K_{\text{RF}}^{\text{B}}=4$. The ground/airborne antennas are spaced by half wavelength. The complex-valued gain $h_u^t$ of each NLoS path is generated following the i.i.d. complex Gaussian distribution with zero mean and unit variance. Under the Cartesian coordinate system, the locations of the LGS and the IGS are set at $(0, 0, 0)$ meters and $(100, 100, 0)$ meters, respectively. The AC is at the initial location of $(0, 0, 8000)$ meters. For the simulation platform, the proposed schemes are implemented using Python with the Pytorch library and an NVIDIA GeForce RTX 3060 GPU processor. 

To avoid the challenging training samples acquisition and online training process, we consider to train the proposed 3DConvTransformer and KDDD-IRN networks offline~\cite{9127834,9849060,9130130}. The simulated dataset is generated based on the classic A2G multi-path channel model provided in~\cite{994803} and the random impulse interference model given in~\cite{4702777} where the emission interval of the impulse signals follows the Poisson distribution. The feasibility of this approach and the robustness of the pretrained networks will be verified in Section V-B and V-C, for demonstrating that the networks pretrained by a simulated dataset can work in practical scenarios with negligible performance loss. Note that, for the training and evaluation of the proposed 3DConvTransformer and KDDD-IRN, we divide the dataset into the training set, the validation set and the test set. The different datasets are generated independently to guarantee the effectiveness of the evaluation. 

\subsubsection{Training of the proposed 3DConvTransformer}

In the offline training phase, we employ the end-to-end learning where all the weights are jointly optimized for both the 3D-CNN and the transformer model. The number\footnote{Note that the $P$ and $L$ values setup can affect the prediction performance of the proposed 3DConvTransformer. Specifically, $P$ determines how much historical IPN information to capture, while $L$ indicates the period over which the IPN needs to be re-estimated. Considering the tradeoff between the training complexity and the prediction accuracy, we set $P=25$ and $L=5$ for demonstration, while the proposed 3DConvTransformer can also be re-trained under other $P$ and $L$ values.} of past IPN observations fed into the 3DConvTransformer is set to $P = 25$, while the temporal prediction range is set to $L = 5$. The AC velocity is generated uniformly following the distribution of $\mathcal{U}[\text{300km/h},\text{600km/h}]$. In the $a$-th layer of the 3D-CNN, we set the convolutional kernel size $K$ to $3\times3\times3$. In addition, the past IPN observation errors are generated randomly following the Gaussian distribution of mean $0$ and variance $\sigma_{\text{e}}^2 = 10$. To simplify the notation, we define $\rho$ as $\rho \text{[dB]} =10\lg\left(\sigma_{\text{e}}^2/1\right)$.  
A learning rate of 0.0001 is adopted with the Adam optimizer. 
\subsubsection{Training of the proposed KDDD-IRN}
We set the size of the training set, the validation set and the test set as $1280$, $128$, and $128$, respectively. In the offline training phase, the learning rate is set to $0.001$, the signal-to-noise power ratio (SNR), which is defined as $\text{SNR}=\frac{1}{\sigma_n^{2}}$ with the normalized LGS transmit power, is set to $8$~dB, and the signal-to-interference power ratio (SIR), which is defined as $\text{SIR}=\frac{1}{\sigma_i^{2}}$ with the normalized LGS transmit power and the IGS transmit power of $\sigma_i^2$, is set to $-3.8$~dB. It is worth noting that, although the KDDD-IRN is trained with the specific SNR and SIR values, the well-trained KDDD-IRN can achieve satisfactory online performance with different SNR and SIR values thanks to the generalization ability. 
\subsection{Performance of the Proposed 3DConvTransformer}
\begin{figure}
\centering
\subfigure[]{\label{CAE1}
\includegraphics[scale=0.4]{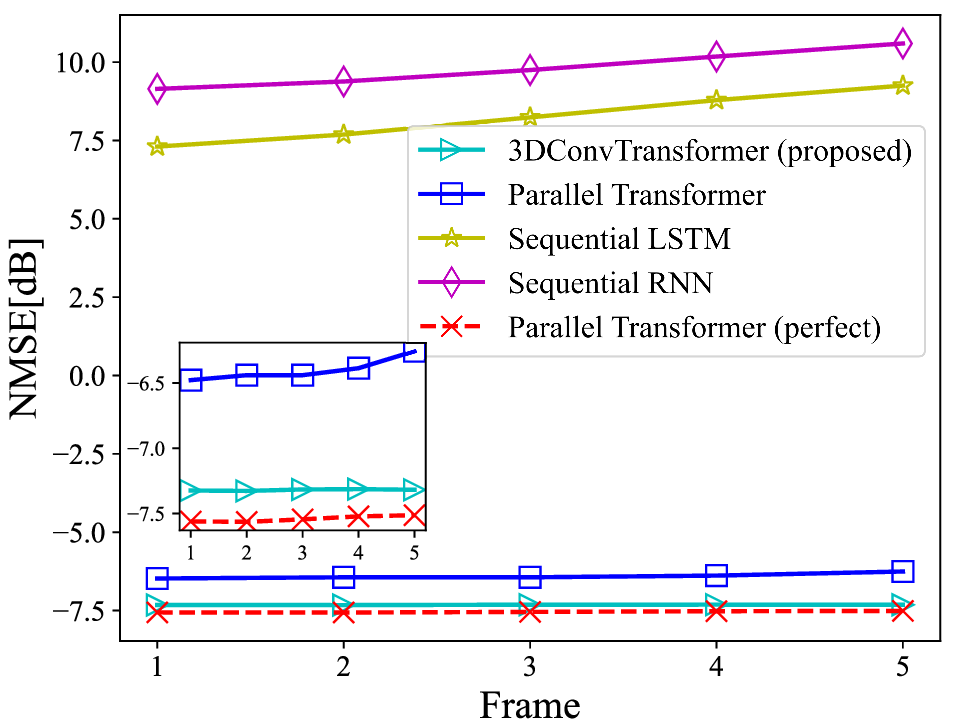}}
\subfigure[]{\label{CAE2}
\includegraphics[scale=0.4]{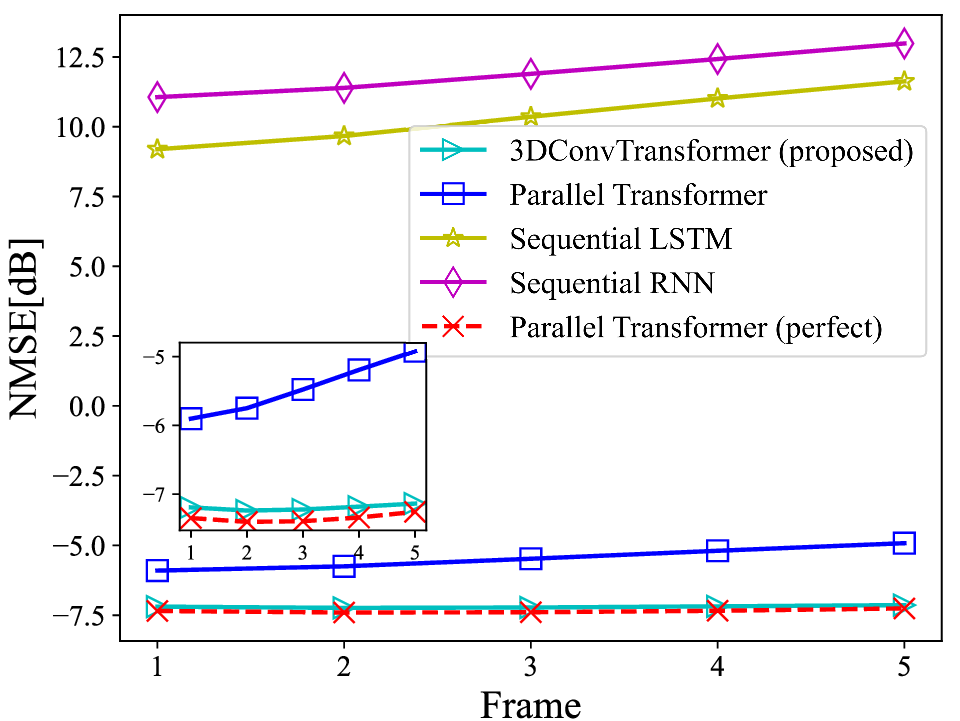}}
\caption{The NMSE performance versus frame with $\rho=$ 10 dB. (a) AC velocity $v=300$~km/h; (b) AC velocity $v=600$~km/h.}\label{fig4}
\end{figure}
To demonstrate the merits of the proposed 3DConvTransformer for IPN covariance matrices prediction, we choose the following benchmarks for comparison:
\begin{itemize}
    \item \textbf{Sequential LSTM}: the re-design of the LSTM proposed in~\cite{9676455} for IPN covariance matrices prediction in a sequential manner;
    \item \textbf{Sequential RNN}: the re-design of the RNN proposed in~\cite{9044427} for IPN covariance matrices prediction in a sequential manner;
    \item \textbf{Parallel transformer}: The transformer model without IPN covariance denoising operations;
    \item \textbf{Parallel transformer (perfect)}: The transformer fed with perfect IPN observations, i.e., $\rho$ $=$ $-\infty$, which provides the upper-bound performance.
\end{itemize}

We first summarize the computational complexity of the proposed 3DConvTransformer and the benchmarks in Table \ref{tab:complexity_table}. According to~\cite{9044427}, the complexity of the sequential RNN is given by $\mathcal{O}\left(F_{\text{R}}(P+L)K_\mathrm{A}^4X^2\right)$, where $F_{\text{R}}$ is the number of the hidden layers. For the sequential LSTM, it is composed of four core computing units: the forget gate, the input gate, the output gate, and the candidate hidden state~\cite{9676455}. As such, the total computational cost is four times that of the sequential RNN, i.e., $\mathcal{O}\left(4F_{\text{L}}(P+L)K_\mathrm{A}^4X^2\right)$, with $F_{\text{L}}$ being the number of hidden layers in LSTM. Furthermore, we utilize the number of floating point operations (FLOPs) that count the total number of computations, i.e., addition, subtraction, multiplication, division, exponentiation and square root, to provide the numerical values of the computational complexity~\cite{10463115}. Specifically, the total number of FLOPs for 3DConvTransformer, parallel transformer, sequential LSTM, and sequential RNN are around $11.02$ Giga FLOPs (GFLOPs), $8.77$ GFLOPs, $3.18$ GFLOPs, and $1.34$ GFLOPs, respectively. Observe that the proposed model exhibits the highest number of FLOPs compared to the benchmarks, primarily due to complex 3D convolution operations and attention mechanisms. Despite the higher computational complexity, the proposed algorithm shows a significant improvement in the predictive performance as demonstrated in the following.
\begin{table*}
\centering
\caption{Computational complexity of different schemes.}
\label{tab:complexity_table}
\small
\renewcommand{\arraystretch}{1.2}
\begin{tabular}{|>{\centering\arraybackslash}m{3cm}|>{\centering\arraybackslash}m{10cm}|>{\centering\arraybackslash}m{3cm}|}
    \hline
    \textbf{Schemes} & \textbf{Complexity}& \textbf{Number of FLOPs} \\ \hline
    3DConvTransformer &
    $\mathcal{O}\left(KK_\mathrm{A}^2X \left(P^2+\sum_{a=1}^A 
C_{a}^{\text{in}}C_{a}^{\text{out}}\right)\right)+\mathcal{O}\left(K_\mathrm{A}^4X^2+\left(G+P+L\right)K_\mathrm{A}^2X\right)$&
    $11,018,616,832$ \\ \hline
    Parallel Transformer &$\mathcal{O}\left(K_\mathrm{A}^4X^2+\left(G+P+L\right)K_\mathrm{A}^2X\right)$&
$8,774,483,968$\\ \hline
    Sequential LSTM &$\mathcal{O}\left(4F_{\text{L}}(P+L)K_\mathrm{A}^4X^2\right)$&
$3,177,709,568$\\ \hline
    Sequential RNN &$\mathcal{O}\left(F_{\text{R}}(P+L)K_\mathrm{A}^4X^2\right)$&
$1,340,833,792$ \\ \hline
\end{tabular}
\end{table*}

Fig.~\ref{fig4} depicts the NMSE of the predicted IPN covariance matrices in future frames with $\rho=10$~dB, under AC velocity of $v=300$~km/h and $v=600$~km/h in Fig.~\ref{CAE1} and Fig.~\ref{CAE2}, respectively. One can observe that the proposed 3DConvTransformer outperforms the parallel transformer by around $1$~dB and achieves near-optimal performance with small gap in comparison with the ``parallel transformer (perfect)", which demonstrates the superiority of 3DConvTransformer in eliminating the initial IPN sampling errors. It can also be observed that the NMSE achieved by LSTM and RNN increases rapidly with the increase of the temporal range. This is caused by error propagation of the sequential operation, which leads to the degradation of the prediction accuracy. In contrast, the performance of 3DConvTransformer stays stable for the future $5$ frames thanks to the parallel signal processing. For example, in the $5$-th frame, the 3DConvTransformer can achieve $15$~dB and $19$~dB better NMSE performance with $v=300$~km/h and $v=600$~km/h, respectively.

Comparing Fig.~\ref{CAE1} and Fig.~\ref{CAE2}, on the one hand, we can find that, with the increase of the AC velocity, both the LSTM and RNN achieve visibly worse performance. This is due to the fact that, as the A2G coherence time gets shorter with higher AC velocity, the early parts of the historical observations become outdated for the prediction of the future ones. On the other hand, the transformer-based prediction schemes stay robust in higher mobility scenarios. This advantage comes from the attention mechanism which helps to put higher weights on the latter parts of the historical observations to improve the prediction accuracy.
\begin{figure}
\centering
\includegraphics[scale=0.4]{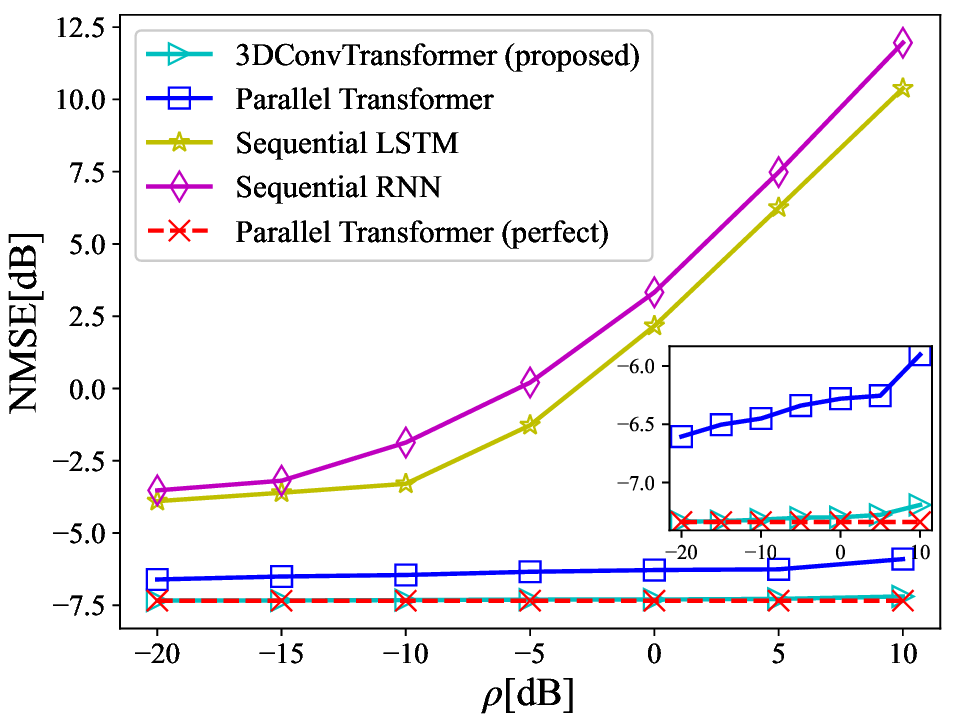}
\caption{{The NMSE performance versus $\rho$ with $v=600$km/h.}}\label{CAE3}
\end{figure}

\begin{figure}
    \centering
\includegraphics[scale=0.4]{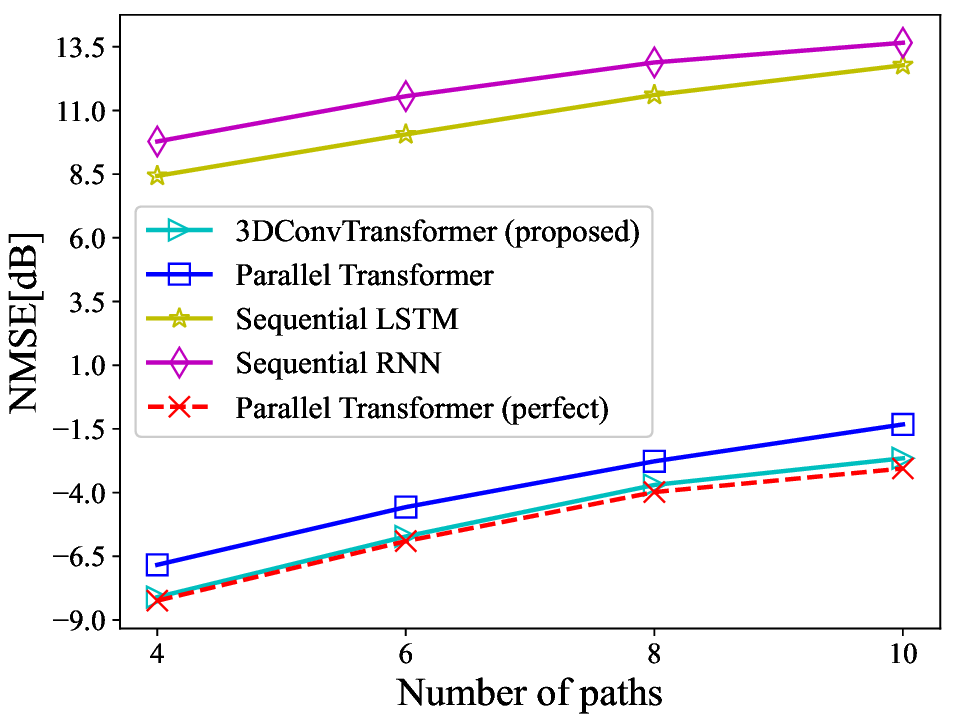}
    \caption{The NMSE performance versus the number of channel paths with $v=600$km/h.}\label{CAE4}
\end{figure}

In Fig.~\ref{CAE3}, we further evaluate the performance of the 3DConvTransformer-based prediction scheme by analyzing the variation of the NMSE with respect to $\rho$. One can first observe that the proposed network trained with $\rho=10$~dB shows its robustness with respect to different $\rho$ values. This performance is due to the fact that, the network can learn the effective mapping between the imperfect past IPN covariance matrices and the future predicted ones, where the influence of the IPN covariances errors is reduced to the minimal. Moreover, observe that the NMSE achieved by the LSTM and RNN increases appreciably with the increase of $\rho$, which is expected as the error propagation problem of the sequential operations is exacerbated with higher initial sampling error values. On the contrary, the parallel prediction schemes show stability with different $\rho$ values. The 3DConvTransformer is shown to outperform the transformer, revealing the effectiveness of 3DConvTransformer for improving prediction accuracy. 

Fig.~\ref{CAE4} compares the NMSE achieved by different schemes versus the number of channel paths, i.e. $(U+1)$. The proposed network and the benchmarks are trained using the data samples with varying $(U+1)\in\left\{3,5,7,9\right\}$ and tested over the data samples generated with $(U+1)\in\left\{4,6,8,10\right\}$. It can be seen that the proposed 3DConvTransformer shows a robust generalizability to $(U+1)$ and superiority to other schemes. As such, the pre-trained 3DConvTransformer is expected to work in practical scenarios with minimal performance degradation given the mismatch in the number of channel paths.
\subsection{Performance of the Proposed KDDD-IRN}
To illustrate the merits of the proposed KDDD-IRN compared to the conventional model-based or data-driven approaches for interference-robust HBF, we choose the following benchmarks as comparison:
\begin{itemize}
    \item \textbf{CNN-IR}: the re-design of the CNN proposed in~\cite{9763852} for purely data-driven interference-resistant HBF; 
    \item \textbf{FD-IR}: the re-design of the fully-digital beamforming approach proposed in~\cite{974266} by considering the interference to provide the upper-bound performance;
    \item \textbf{AO-IR(5,2)}: The AO approach as introduced in Section IV-A, where the number of outer iterations is set to $2$, while the number of inner iterations for the first and second outer loops are set to $5$ and $2$, respectively;
    \item \textbf{AO-IR(converge)}: The AO-IR runs until convergence, which provides the local-optimal performance achieved by the AO-IR.
\end{itemize}
All of the above benchmarks take the IPN covariance matrices predicted by the proposed 3DConvTransformer as input. 

\begin{figure}
\centering
\subfigure[]{\label{Figure_loss1}
\includegraphics[scale=0.4]{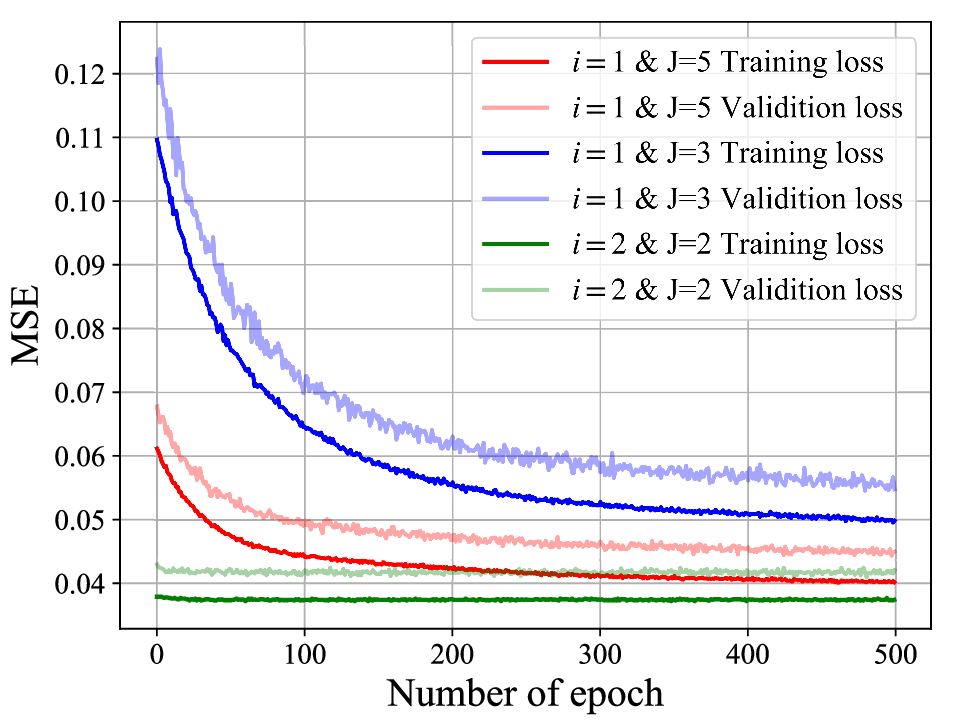}}
\subfigure[]{\label{Figure_loss2}
\includegraphics[scale=0.4]{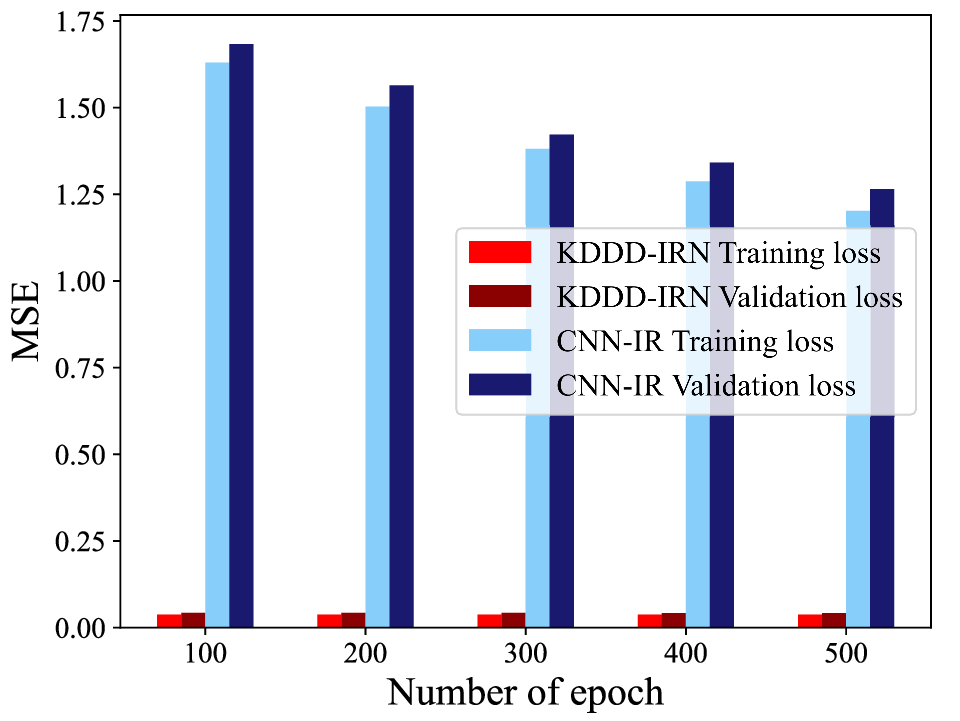}}
\caption{(a) Performance of KDDD-IRN in the training phase; (b) Performance comparison between the KDDD-IRN and CNN-IR in the training phase.}\label{Figure_loss}
\end{figure}
In Fig.~\ref{Figure_loss1}, we demonstrate the learning process of the proposed KDDD-IRN, where the number of layers, i.e., $I$, is set to $2$. As the layer-by-layer training strategy is adopted, the training process of the first and second layers, which are labeled with ``$i=1$" and ``$i=2$", respectively, are both depicted. It is shown that the training loss, i.e., the MSE, converges within around $300$ epochs and stay stable afterwards, which demonstrates the effectiveness of the learning mechanism. Moreover, for the first-layer training, it is observed that the training loss is significantly reduced when the number of gradient descent steps, i.e., $J$, increases, which is expected due to the higher degree of freedom for optimization. For the second-layer training, the output of the well-trained first layer is fed as the initial values for further optimization. As such, the training loss of the second layer gets smaller than that of the first layer. Considering the tradeoff between the complexity and the performance, we set $J$ to $5$ and $2$ for the first and second layers, respectively, (denoted by ``KDDD-IRN(5,2)"), in the rest of this section for further demonstrating the KDDD-IRN performance. We also observe that the KDDD-IRN shows comparable performance over the validation set compared to that over the training set. Since the validation and training sets are generated independently, the generalizability of KDDD-IRN with respect to different input, i.e., $\mathbf{H}_{\text{BA}}^x$ and $\mathbf{R}_x$, is verified. 

To demonstrate the superiority of our proposed KDDD-IRN compared to the purely data-driven methods, we select the CNN-based HBF design proposed in~\cite{9763852} as the benchmark. Note that, to the best of the authors' knowledge, there is no existing learning-based approach for the IR-HBF design. Therefore, we re-designed the CNN network proposed in~\cite{9763852}, which constructed the mapping from the CSI to the HBF matrices with the aim of sum rate maximization under the environment without interference. Specifically, we modified the loss function to the MSE while feeding the IPN covariance matrix as the input, for fair comparison. In Fig.~\ref{Figure_loss2}, we present the training and validation loss of both CNN-IR and KDDD-IRN using the same training and validation dataset, with the sizes of $1280$ and $128$, respectively. One can observe that KDDD-IRN achieves significantly lower training/validation loss compared to CNN-IR. This is expected since KDDD-IRN has a much simpler network architecture with a reduction in the number of training parameters, and thus requires less training data. In fact, observing Fig.~\ref{Figure_loss1}, it is shown that the KDDD-IRN can be well trained with the training data set of $1280$, while it was demonstrated in~\cite{9763852} that the CNN-IR requires at least $7000$ training samples. As such, the inadequate training data leads to the performance deterioration of CNN-IR due to the underfitting. One can also observe that the CNN-IR can not converge after $500$ epochs, while the KDDD-IRN is shown to converge within around $300$ epochs in Fig.~\ref{Figure_loss1}. This can also be explained by the simplified processing and less training parameters in KDDD-IRN, which results in shorter training time. These observations reveal the appealing advantage of KDDD-IRN in practical scenarios where both the data samples and training time are limited.

\begin{figure}
  \centering
    \begin{minipage}[b]{\linewidth}
      \subfigure[]{\label{figa}
\includegraphics[width=0.48\linewidth]{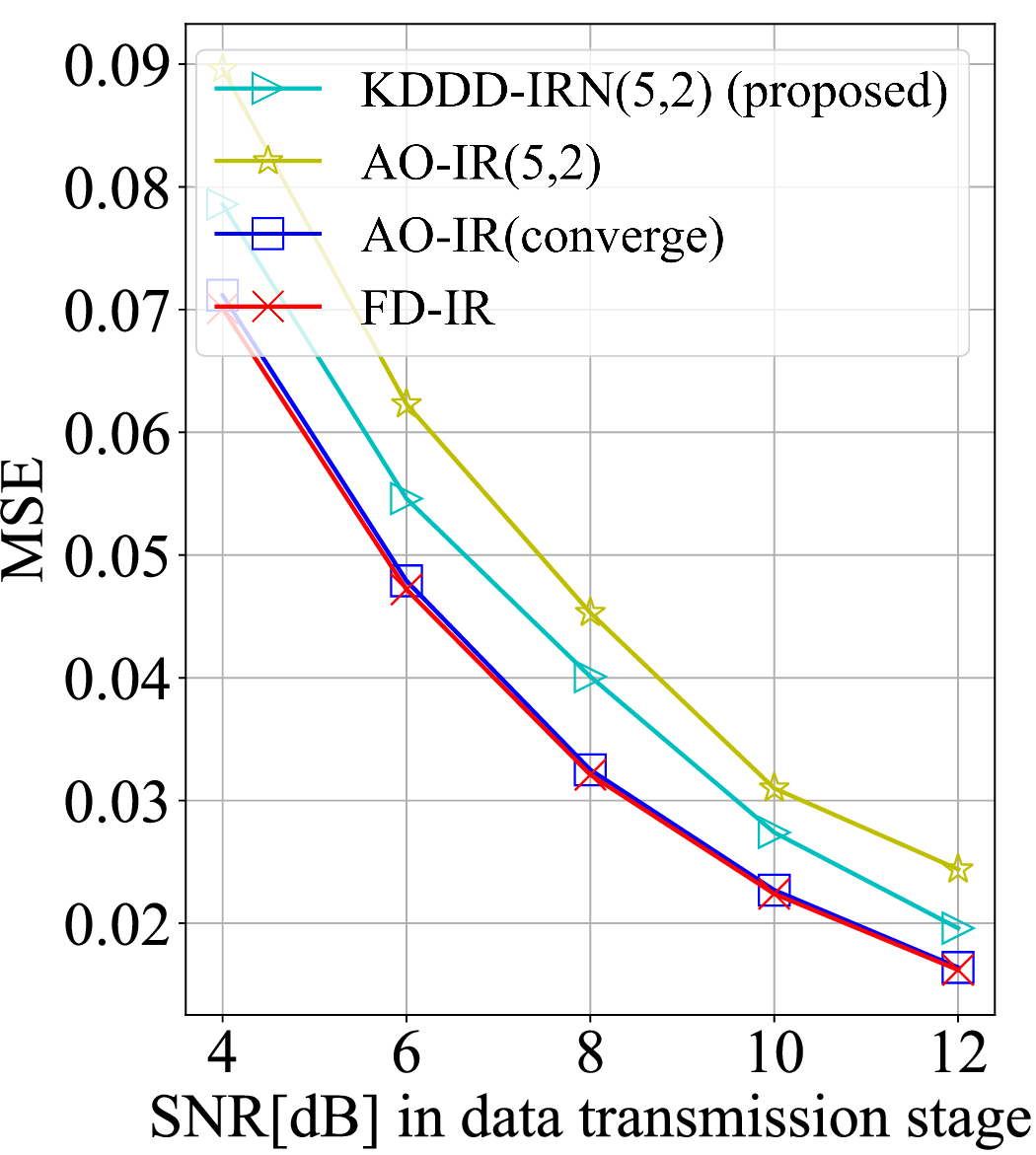}
      }\hspace{-3mm}
      \subfigure[]{\label{figb}
\includegraphics[width=0.48\linewidth]{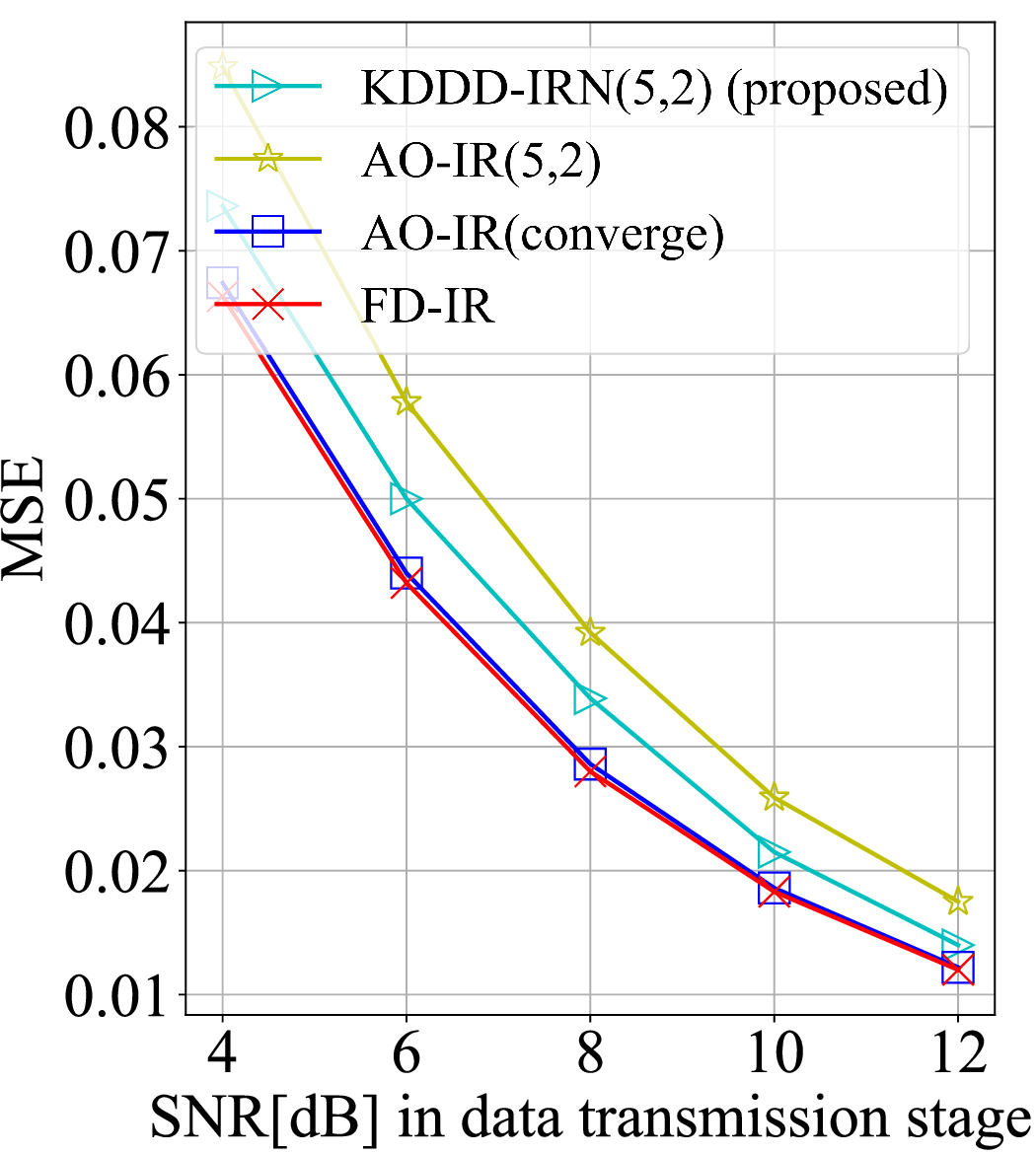}
      }
    \end{minipage}
    \caption{The achievable MSE versus SNR with SIR $=$ $-3.8$ dB. (a) $\rho = -20$ dB; (b) $\rho = -30$ dB.}\label{fig:Figure_1}
\end{figure}

Fig.~\ref{fig:Figure_1} depicts the MSE between the received signal and the transmitted desired signal obtained by different interference-resistant beamforming schemes versus the SNR, where the SIR is set to $-3.8$ dB, while $\rho$ is set to $-20$~dB and $-30$~dB in Fig.~\ref{figa} and Fig.~\ref{figb}, respectively. Observe that the KDDD-IRN(5,2) outperforms AO-IR(5,2) under different SNR values, which is expected as the KDDD approach brings in performance gain compared to the conventional model-based optimization thanks to the offline data training. For example, for the HBF with SNR = $6$~dB and $\rho = -20$~dB, the KDDD-IRN(5,2) can reduce the MSE achieved by the AO-IR(5,2) by around $15\%$, and achieves nearly $87.8\%$ of the upper-bound performance provided by the FD-IR. Moreover, since the AO-IR solves the HBF problem with deterministic calculations, the superiority of the KDDD-IRN over the AO-IR(5,2) at all SNR values implies the strong robustness of the former to different SNRs. This is expected, as the KDDD-IRN replicates the conventional model-based algorithm where the effect of the noise power is taken into consideration in the calculation process. As such, the performance of the KDDD-IRN can be guaranteed given the mismatch between the SNRs in the simulated and the realistic environment.
\begin{figure}
  \centering
    \begin{minipage}[b]{\linewidth}
      \subfigure[]{\label{figaaa}
        \includegraphics[width=0.48\linewidth]{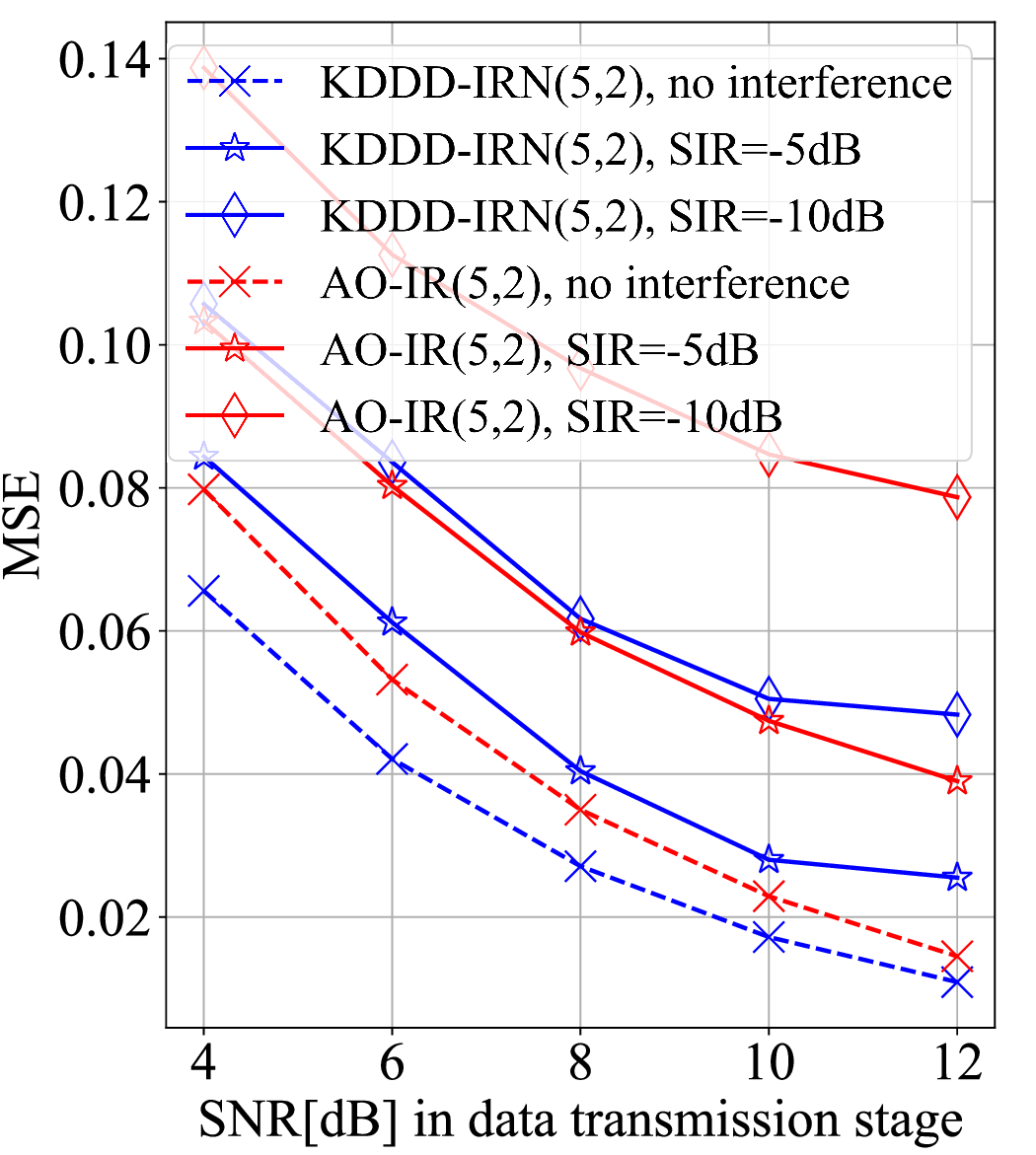}
      }\hspace{-3mm}
      \subfigure[]{\label{figbbb}
        \includegraphics[width=0.48\linewidth]{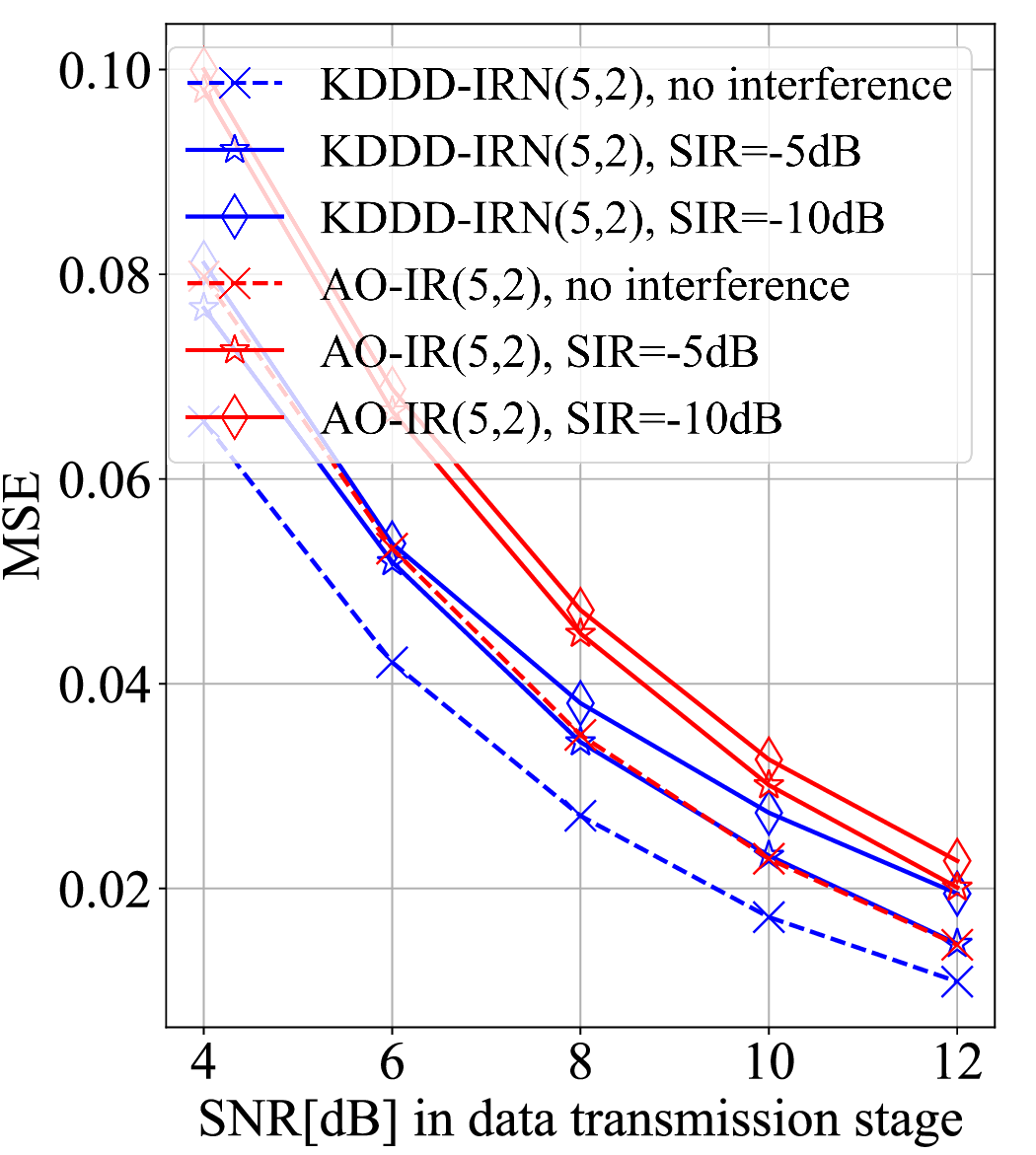}
      }
    \end{minipage}
    \caption{The achievable MSE versus SNR with different SIR. The red dashed line and the blue dashed line represent the performance upper bounds when no interference exists. (a) $\rho = -20$ dB; (b) $\rho = -30$~dB.}\label{Figure_3}.
  \end{figure} 
  
\begin{figure}
  \centering
    \begin{minipage}[b]{\linewidth}
      \subfigure[]{\label{figaa}
        \includegraphics[width=0.48\linewidth]{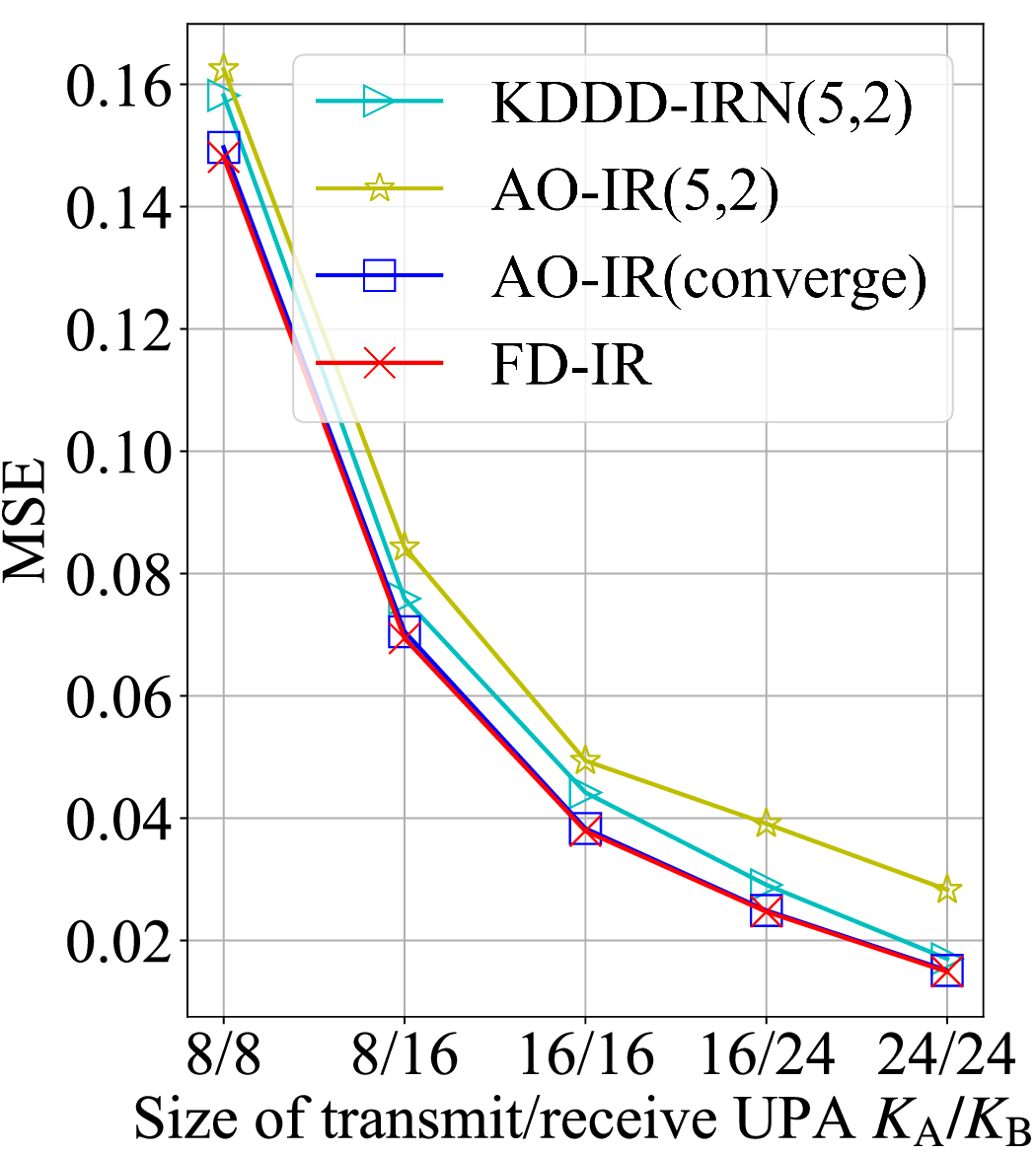}
      }\hspace{-3mm}
      \subfigure[]{\label{figbb}
        \includegraphics[width=0.48\linewidth]{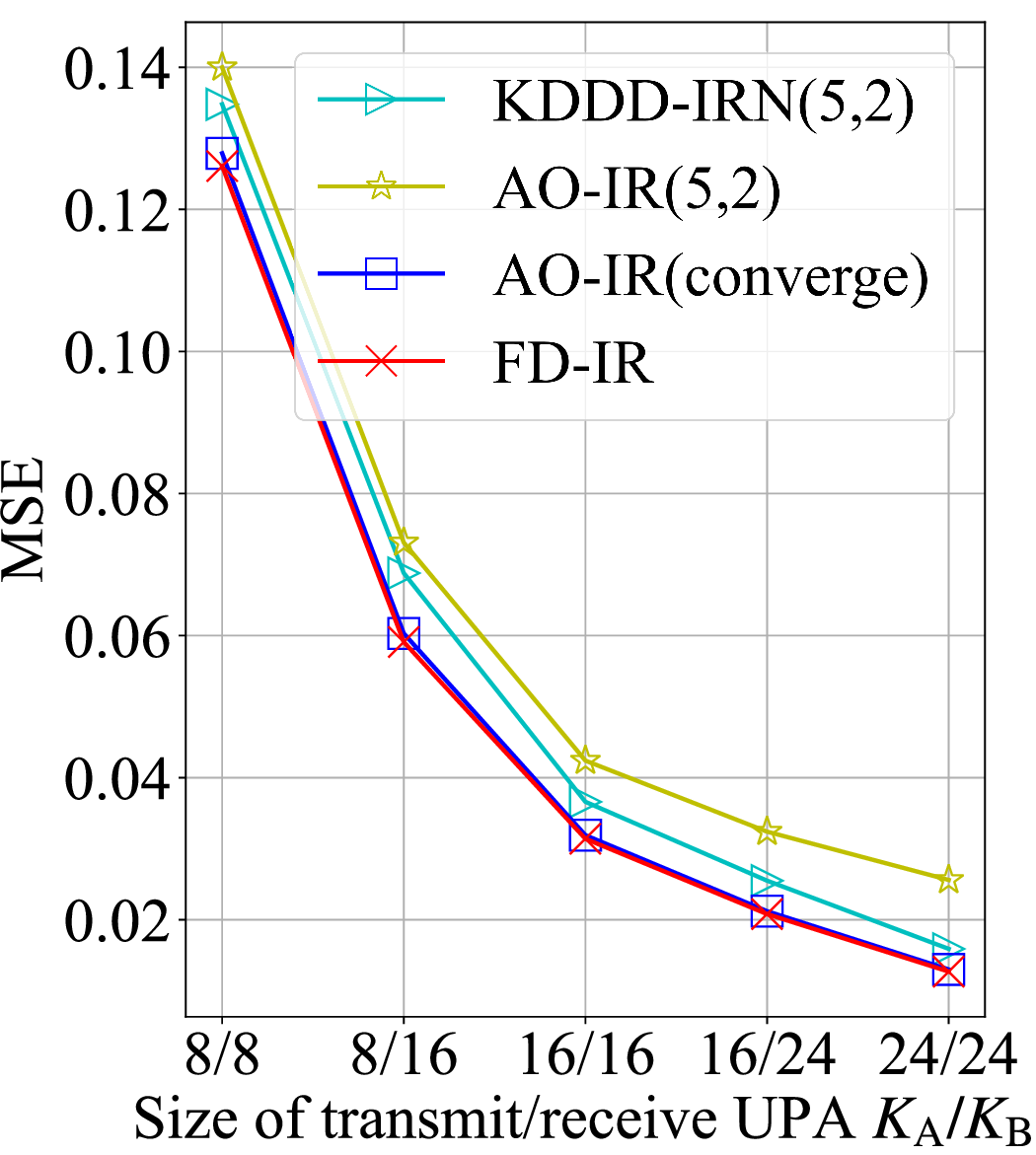}
      }
    \end{minipage}
    \caption{The achievable MSE versus antenna configuration with SIR $=$ $-3.8$ dB and SNR $=$ $8$ dB. (a) $\rho = -20$ dB; (b) $\rho = -30$~dB.}\label{Figure_2}.
  \end{figure}

The performance comparison between the AO-IR and the KDDD-IRN under different SIR values is demonstrated in Fig.~\ref{Figure_3}. It is evident to see from Fig.~\ref{figaaa} that, the superiority of the KDDD-IRN over the AO-IR becomes more obvious with higher interference power. For example, the KDDD-IRN reduces the MSE by around $34\% $ and $57\% $ compared to the AO-IR with SIR $= -5$~dB and SIR $= -10$~dB, respectively. This is a very encouraging appearance which verifies that the KDDD-IRN is robust against random interference with strong interference power. Additionally, one can observe from Fig.~\ref{figbbb} that the performance gap between the cases of SIR $= -5$~dB and SIR $= -10$~dB is quite small for both the KDDD-IRN and the AO-IR. This reveals that the interference-resistant schemes can achieve satisfying performance even with high interference power as long as the estimation error of the JPN covariance matrix is small.

Fig.~\ref{Figure_2} depicts the MSE versus antenna configurations with SIR $=-3.8$~dB and SNR $=8$~dB, where $\rho$ is set as $-20$~dB and $-30$ dB in Fig.~\ref{figaa} and Fig.~\ref{figbb}, respectively. It is observed that the MSE decreases with the increase of the number of transmit/receive antennas, which is brought by the boosting of the beam gain in the desired direction as well as the null steering in the interference direction. Furthermore, the KDDD-IRN(5,2) surpasses the AO-IR(5,2) with larger gap and approaches AO-IR(converge) as $K_\text{A}/K_\text{B}$ increase. One important reason is that, as the problem complexity becomes higher with the increase of the number of transmit/receive antennas, the AO-IR generally requires
more iterations for convergence, thereby leading to poorer performance under limited number of outer/inner iterations. On the contrary, KDDD-IRN shows its robustness against different antenna configurations, given that the network is intelligently trained to boost its performance under the predefined iterative structure. For example, when $\rho$ is set to $-20$~dB, KDDD-IRN(5,2) can reduce the MSE by around $10\%$ and $37\%$ compared to the AO-IR(5,2) with $K_\text{A}/K_\text{B}=16/16$ and $K_\text{A}/K_\text{B}=24/24$, respectively. Besides, KDDD-IRN(5,2) can achieve nearly $97\%$ of the upper-bound performance achieved by the FD-IR with $K_\text{A}/K_\text{B}=24/24$, which demonstrates the superiority of the proposed KDDD-IRN for solving more complicated HBF problems.  

\begin{figure}
    \centering
    \includegraphics[scale=0.4]{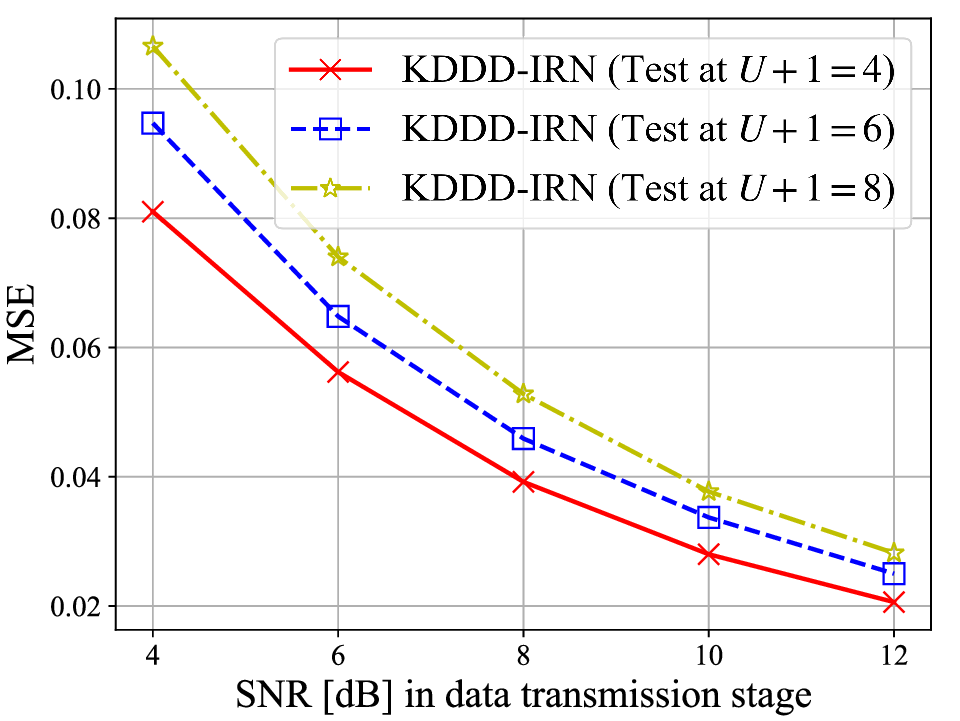}
    \caption{The MSE performance of the KDDD-IRN tested with different numbers of channel paths, where SIR=$-3.8$~dB, $\rho=-20$~dB.}
    \label{fig:KDDD-IRN-path}
\end{figure}

Fig.~\ref{fig:KDDD-IRN-path} demonstrates the robustness and generalizability of the KDDD-IRN in various communications environments that involve changes in the number of channel paths. In the figure, the KDDD-IRN  is pretrained with $(U+1) = \left\{3,5,7,9\right\}$ and tested over the data samples generated with $(U+1) = \left\{4,6,8\right\}$ channel paths. As can be observed, the proposed network performs well at all $(U+1)$ values, which shows its robustness and demonstrates that the KDDD-IRN can be applied to realistic scenarios with different numbers of channel paths, without having to retrain the entire network structure.
      
\begin{figure}
	\centering
 \includegraphics[scale=0.4]{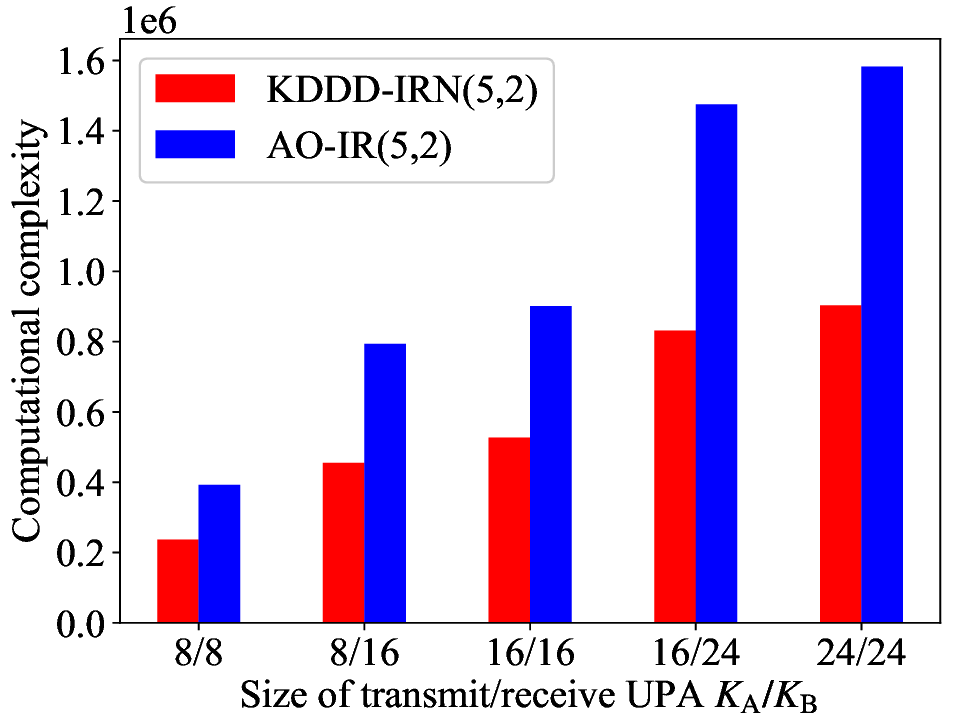}	
 \caption{Interference-resistant HBF complexity with SIR $=$ $-3.8$~dB, SNR $=$~8 dB, and $\rho = -20$ dB.}\label{complexity}
\end{figure}

Fig.~\ref{complexity} depicts the computational complexity of KDDD-IRN and AO-IR versus antenna configuration with SIR $=$ $-3.8$~dB, SNR $=$ 8~dB and $\rho = -20$~dB. Note here that the computational complexity of the FD-IR is not provided. Since the FD-IR is for the fully-digital beamforming design, it can not provide fair comparison between the KDDD-IRN and the FD-IR on the computational complexity as they are aiming for solving different problems.
It is shown that the computational complexity for both AO-IR and KDDD-IRN increases with the increase of $K_\text{A}/K_\text{B}$, which is intuitive as the dimensions of the optimization variables get larger.
Furthermore, observe that when the numbers of outer/inner iterations are fixed, KDDD-IRN can reduce the computational complexity by around $40\%-45\%$ compared to the conventional AO-IR approach. This advantage comes from the fact that KDDD-IRN intelligently learns gradient descent step sizes offline and adopts the predefined values online, which replaces the complicated backtracking line search, thereby significantly reducing the complexity. Combining the results of Fig.~\ref{complexity} and Fig.~\ref{Figure_2}, we can find that the KDDD-IRN can achieve better performance than AO-IR with much lower complexity. 
\section{Conclusions}
In this paper, a time-efficient framework was proposed for interference-robust broadband MIMO communications under rapidly-varying channels. To obtain the unknown interference statistics, we designed an IST module where the IPN covariance matrices were predicted in parallel by exploiting the time- and spatial-domain correlations. Benefiting from the data training, the IST module can realize fast and accurate prediction with even zero pilot overhead, which well suits the rapidly-varying channels. Moreover, based on the predicted IPN covariance matrices, we further investigated the low-complexity IR-HBF module, where the KDDD-IRN was designed to incorporate both the prior knowledge of AO operations and the data training, so as to construct the interpretable mapping from the interference statistics to the beamforming weights matrices. Our simulation results demonstrated the robustness of the proposed framework against random interference even with strong interference powers. For future study, the interference-resistant beamforming in the high-mobility multi-user MIMO communications needs to be investigated, where both the external interference and the inter-user interference should be suppressed efficiently with low complexity.

\bibliographystyle{IEEEtran}
\bibliography{mybib}
\end{document}